\documentclass{LMCS}
\usepackage{graphicx}
\usepackage{latexsym}
\usepackage{amsfonts}
\usepackage{amsthm}
\usepackage{epic}
\usepackage[all]{xy}
\usepackage{enumerate,hyperref}

\newif\iflncs\lncsfalse

\newcommand{\nats}{{\mathbb N}}
\newcommand{\deq}{{\downarrow}{\raisebox{1ex}{\rmfamily\hspace{-1.3ex}\scriptsize=}}}
\newcommand{\deqsm}{\downarrow{}^{\hspace{-1.1ex}=}}

\newcommand{\val}{\mathchoice{\mbox{\it Val}}{\mbox{\it Val}}%
{\mbox{\scriptsize\it Val}}{\mbox{\scriptsize\it Val}}}
\newcommand{\parm}{\mbox{\it Var}}
\newcommand{\maxval}[2]{{\mbox{\sc MAXVAL}_{#1}{#2}}}
\newcommand{\minval}[2]{{\mbox{\sc MINVAL}_{#1}{#2}}}
\newcommand{\atseret}[1]{{#1!\,}}
\newcommand{\sep}{\ |\ } 

\iflncs
\spnewtheorem{obs}[theorem]{Observation}{\bfseries}{\rmfamily}
\spnewtheorem{construction}[theorem]{Construction}{\bfseries}{\rmfamily}
\else
\theoremstyle{plain}\newtheorem{claim}[thm]{Claim}
\theoremstyle{plain}
\fi

\newcommand{\com}{\newcommand}
\com{\bthm}{\begin{thm}}
\com{\ethm}{\end{thm}}
\com{\bdfn}{\begin{defi}}
\com{\edfn}{\end{defi}}
\com{\blem}{\begin{lem}}
\com{\elem}{\end{lem}}

\com{\bprf}{\proof}
\com{\eprf}{\qed}

\newcommand{\bi}{\begin{itemize}}
\newcommand{\ei}{\end{itemize}}
\newcommand{\be}{\begin{enumerate}}
\newcommand{\ee}{\end{enumerate}}
\newcommand{\ba}{\begin{array}}
\newcommand{\ea}{\end{array}}
\com{\pgt}[1]{{\tt #1}}
\renewcommand{\vec}[1]{\mathbf{#1}}

\com{\fl}{\noindent}
\com{\hair}{\hspace{3mm}}
\com{\vair}{\vspace{3mm}}

\com{\ints}{{\mathbb Z}}
\com{\nat}{{\mathbb N}}
\com{\deqarrow}{{\downarrow}%
  {\raisebox{1ex}{\rmfamily\hspace{-1.3ex}\scriptsize=}}}

\renewcommand{\Im}{\mathrm{Im}}
\newcommand{\nxt}{\mbox{\rm\upshape next}}

\def\doframeit#1{\vbox{%
  \hrule height\fboxrule
    \hbox{%
      \vrule width\fboxrule \kern\fboxsep
      \vbox{\kern\fboxsep #1\kern\fboxsep }%
      \kern\fboxsep \vrule width\fboxrule }%
    \hrule height\fboxrule }}

\def\frameit{\smallskip \advance \linewidth by -7.5pt \setbox0=\vbox \bgroup
\strut \ignorespaces }

\def\endframeit{\ifhmode \par \nointerlineskip \fi \egroup
\doframeit{\box0}}

\newenvironment{fig0}[3]
{
\xdef\fighack{\noexpand\caption{#1}\noexpand\label{#3}}
\begin{figure*}[#2]
\begin{frameit}%
\hspace*{3mm}\begin{minipage}{0.95\textwidth}}{\end{minipage}%
\fighack%
\end{frameit}%
\end{figure*}}

\newenvironment{fig1}[3]
{
\xdef\fighack{\noexpand\caption{#1}\noexpand\label{#3}}
\begin{figure*}[#2]
\hspace*{3mm}\begin{minipage}{0.95\textwidth}}{\end{minipage}%
\fighack%
\end{figure*}}
\newcommand{\programenvironment}{\programmode%
	\def\par{\leavevmode\endgraf}\obeylines\nobreak%
	\programmode}

\newcommand{\programmode}{\tt%   Is: \LARGE\tt, PS 1992-09-23
	\catcode`\_=12 \catcode`\?=12 \catcode`\.=12 \catcode`\,=12
	\catcode`\;=12 \catcode`\:=12 \catcode`\@=12 \catcode`\~=12
         \catcode`\#=12 \catcode`\&=12      % Added by PS 1991-06-17
	\obeyspaces\frenchspacing}%

\newenvironment{programintext}{\programenvironment}{}

\newenvironment{myprogram}{\setlength{\partopsep}{0mm}\setlength{\topsep}{0mm}
	\begin{trivlist}\item[]
	\hspace*{2em}  %according to TOPLAS formatting instructions
	\begin{minipage}{0.9\textwidth}
	\vspace{1mm}
	\begin{programintext}
	}{\end{programintext}
	\vspace{\belowdisplayskip}
	\end{minipage}
	\end{trivlist}
	\noindent}

\com{\bp}{\begin{myprogram}}
\com{\ep}{\end{myprogram}}

\def\doi{5 (2:8) 2009}
\lmcsheading%
{\doi}
{1--29}
{}
{}
{Nov.~13, 2008}
{May~\phantom{.}25, 2009}
{}   

\begin{document}

\title{Ranking Functions for Size-Change Termination II}

\author[A.~M.~Ben-Amram]{Amir M. Ben-Amram\rsuper a}
\address{{\lsuper a}School of Computer Science, The Tel-Aviv Academic College, Israel}	%required
\email{benamram.amir@gmail.com}  %optional

\author[C.~S.~Lee]{Chin Soon Lee\rsuper b}
\address{{\lsuper b}Singapore}	%required
\email{cslee@asteria.com.sg}  %optional


\begin{abstract}

Size-Change Termination is an increasingly-popular technique for verifying
program termination.
These termination proofs are deduced from an abstract representation
of the program in the form of \emph{size-change graphs}.

We present algorithms that, for certain classes of size-change graphs,
deduce a \emph{global ranking function}: an expression that ranks program
states, and decreases on every transition.
A ranking function serves as a witness for
a termination proof, and is therefore interesting for program
certification.  The particular form of the ranking expressions that
represent SCT termination proofs sheds light on the 
scope of the proof method. The complexity of the expressions
is also interesting, both practicaly and theoretically.

While deducing ranking functions from size-change graphs has already
been shown possible,
the constructions in this paper
are simpler and more transparent than previously known.
They improve the upper bound on the size of the ranking expression
from triply exponential down to singly exponential  (for certain classes of instances).
We claim that this result is, in some sense,
optimal.  To this end, we introduce a framework for lower bounds on the complexity
of ranking expressions and prove exponential lower bounds.

\end{abstract}

\maketitle

\section{Introduction}

Automatic termination analysis is a rapidly growing field; it represents
exciting progress in the theory and application of program analysis.
Two widely-cited foundational publications are
Podelski and Rybalchenko~\cite{PR-LICS04} and Lee, Jones and Ben-Amram%
~\cite{leejonesbenamram01}. The former promoted the use of \emph{local
ranking functions} (or relations) for termination proofs; the latter presented
the Size-Change Termination (SCT) framework. SCT is, in essence, a class
of abstract programs (in other words, transition systems), that happen
to be conveniently represented as graphs (known as size-change graphs).
These abstract programs
can be used to safely approximate the semantics of an actual program,
while capturing invariants that are crucial to a termination proof.

A \emph{global ranking function} is a function of program states that decreases
towards its lower bound in every program transition, thus providing a direct
witness to termination. It is a folklore theorem that a program terminates
if and only if such a function exists.
But global ranking functions
can be complex and difficult to find, even for simple kinds of programs
(for example, the programs that we will consider in this work).
Both~\cite{PR-LICS04} and~\cite{leejonesbenamram01} circumvent the construction
of a global ranking function. In fact, they are closely related:
as clearly explained in~\cite{Codish-et-al:05},  SCT termination proofs
fit the local framework of \cite{PR-LICS04}.

The problem addressed in this work is that of deriving an
explicit expression for a
global ranking function for a given SCT instance
(despite the fact that a termination proof can be obtained without it).
Why are global ranking functions interesting?
Firstly, for theoretical understanding of size-change termination :
identifying a class of ranking functions that captures all terminating
instances
provides a clue to the \emph{scope} of the proof method
(what behaviours are captured by the method, what ordinals
may be captured etc.). Effectively constructing a ranking function 
is a challenge because it is not at all obvious how to do it, even when
a function is known to exist; and developing such an algorithm provides new
insights into the subject.
Secondly, for applications:
an explicit ranking expression may
provide an easy-to-verify witness to termination, since verification
only amounts to checking it against every transition.
Such a witness is not provided by the local methods.
As pointed out
by Krauss~\cite{krauss07}, if a global ranking function (of sufficiently simple
form) could be constructed
automatically, it would allow a theorem prover to certify
the termination claim while allowing the tool that searches for the
termination proof to stay outside the trusted (formally verified) code
base.  One can also consider applications to proof-carrying code,
where again the desire is for the proof to be given as a certificate
that is easier to check than to find. Finally, an interesting potential application
to bound the excution time of programs, since a ranking function provides 
a progress measure. However, such application is not immediate, since the
range of our functions is, in general, not the integers (i.e., not the order type $\omega$,
but $\omega^k$ for some $k$).

\begin{fig0}{A gcd program. One ranking function for the loop
is the maximum among the
variables. Another is their sum.}%
{t}{fig-gcd}
\bp
while (x, y > 0)
\ \ if x>y then x := x-y
\ \ \ \ \ else y := y-x
return max(x,y)        
\ep
\end{fig0}

For all these reasons, we are interested in the form and complexity
 of global ranking functions that suffice
for SCT programs, and in algorithms for their automatic construction.
Early publications on SCT
identified several special cases: programs where the maximum, minimum, 
or sum of 
a set of variables
decreases (Figure~\ref{fig-gcd}), programs with a lexicographic descent
(the ubiquitous Ackermann's function), and programs with 
multiset-descent~\cite{TG:05,BL:2006}.
Lee~\cite{Lee:ranking} established for the first time
that a ranking function can be
automatically constructed for \emph{any terminating SCT instance},
always of the following form:
$$\min ( \max S_1, \max S_2, \dots )$$
where $\max S_i$ represents the maximum element among a set $S_i$ of 
\emph{vectors} (tuples) of 
variables and constants, where vectors are
lexicographically ordered.

%\input ex-cls-new.tex
%("Final mystery descent" from Lee:ranking rearranged for new paper)
Let us give an example. Figure~\ref{fig-ex-cls} shows a program,
and the graphs in Figure~\ref{fig-cls-graphs}
represent it in SCT form. The three graphs correspond
to the three alternatives in the loop body, with arcs representing data flow
and heavy arcs representing descent (for precise definitions of size-change graphs, see 
Section~\ref{sec-formalintro}).

\begin{fig0}{A terminating program. Function f is considered
unknown. ``do either'' is nondeterministic.}{t}{fig-ex-cls}
\bp
while (x,y,z > 0)
\ \  do either
\ \ \ \  (x,y,z) := (y-1, y-1, f(x,y,z))
\ \ \ | (x,y,z) := (x-1, x,   f(x,y,z))
\ \ \ | (x,y,z) := (y-1, y,   z-1)
\ep
\end{fig0}

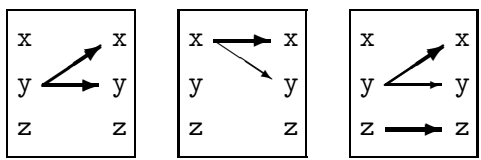
\begin{fig0}{Size-change graphs for the program in Figure~2.}
{t}{fig-cls-graphs}
{\setlength{\unitlength}{0.4pt}
\flushleft\begin{picture}(600,200)(0,60)
\thinlines
	      \put(188,85){\framebox(120,138){}}
	      \put(197,189){\tt x}
	      \put(197,147){\tt y}
	      \put(197,105){\tt z}
	      \thicklines
	      \put(221,153){\vector(3,2){55}}
	      \put(221,152){\vector(3,2){55}}
	      \put(221,153){\vector(1,0){53}}
	      \put(221,152){\vector(1,0){53}}
	      \thinlines
	      \put(287,189){\tt x}
	      \put(287,147){\tt y}
	      \put(287,105){\tt z}
	      
	      \put(350,85){\framebox(120,138){}}
	      \put(359,189){\tt x}
              \put(359,147){\tt y}
              \put(359,105){\tt z}
	      \thicklines
              \put(383,195){\vector(1,0){53}}
	      \put(383,194){\vector(1,0){53}}
	      \thinlines
	      \put(383,195){\vector(3,-2){55}}
	      \put(449,189){\tt x}
              \put(449,147){\tt y}
              \put(449,105){\tt z}

	      \put(512,85){\framebox(120,138){}}
	      \put(521,189){\tt x}
              \put(521,147){\tt y}
              \put(521,105){\tt z}
	      \thicklines
              \put(545,111){\vector(1,0){53}}
	      \put(545,110){\vector(1,0){53}}
	      \put(545,153){\vector(3,2){55}}
	      \put(545,152){\vector(3,2){55}}
	      \thinlines
	      \put(545,153){\vector(1,0){53}}
	      \put(611,189){\tt x}
              \put(611,147){\tt y}
              \put(611,105){\tt z}
\end{picture}} 
\end{fig0}

A ranking function for this program is
$\rho(x,y,z)=\max\{\langle y,0,z\rangle,
\langle x,1,z\rangle\}$
(in this instance the \emph{min} operator is unnecessary;
this could be expected, as Section~\ref{sec-fif} will show).
A straight-forward way to verify this
is to check each graph in turn,
considering each possibility among $y>x$, $y=x$ and $y<x$.
For example, take the first graph. Assume that initially $y>x$.
Then the initial maximum is $\langle y,0,z\rangle$;
since the transition decreases {\tt y}, there is no doubt that $\rho$
descends.

%%%%%%%%%%%%%

\paragraph{Contributions of this work.}
We provide new algorithms to construct ranking functions for a restricted,
but interesting, class of SCT instances: fan-in free (or fan-out free) graphs.
We feel that the new
constructions are far more transparent than the previous one, which
involved a lengthy detour through the determinization of B\"uchi automata.
In contrast, the new ones are based on direct analysis 
of SCT graphs. They employ a technique of
including composite values, such as
tuples or sets of variables, as single data items,
and showing that this simplifies the SCT analysis.
The inclusion of tuples 
reflects the role of lexicographic descent; sets of variables
give rise to descent in \emph{multiset orders} (see Section 2).
Thus the constructions also provide one more example of the
usefulness of multiset orders in termination-related reasoning.
In terms of expression size, we reduce the upper bound from triply-exponential
as in~\cite{Lee:ranking} to singly-exponential. 

An additional contribution of this work is the formulation of a lower-bound
framework and the proof of exponential lower bounds, which imply that our
upper complexity bounds are, in a certain sense, optimal.

\paragraph{Structure of this paper.} The next section provides the formal introduction,
giving necessary definitions. In Section~3 we review some results that we are using from
previous  work.  Then, in Sections~4 and 5 we give the new construction for fan-out
free, then fan-in free, graphs. Section~6 is concerned with lower bounds. In Section~7,
we discuss the connection of our results to a more general
theorem of Blass and Gurevich. Section~8 concludes.

%\tableofcontents

\section{Definitions}  \label{sec-formalintro}

In this section we list the necessary definitions involving SCT,
ranking functions, and their connection. Enough definitions are given to
make the paper formally self-contained and to fix the terminology (which is,
unfortunately, not uniform across SCT-related work). 

\subsection{Program representation}

Let $\val$ be a well-ordered set of data values.

\begin{defi}
A control-flow graph (CFG) is a directed
multigraph $(F,C)$. The nodes are called flow-chart points or just flow-points.
The set of arcs from $f\in F$ to $g\in F$ is denoted $C_{fg}$.

For each $f\in F$,
we have a distinct set of \emph{variables} $\parm(f)$.

One of the nodes, $f_0$,
is \emph{initial} or starting point. All nodes are reachable from $f_0$.
\end{defi}

\noindent Often, we further assume that $f_0$ is also reachable from
all nodes, or equivalently that the graph is strongly connected; it is
well-known that termination analysis can be done one
strongly-connected component at a time.

The variables $\parm(f)$ are supposed to represent
data pertinent to the program state when the program is at point $f$.

To avoid cumbersome notations, we make in this paper the following assumption:

\begin{frameit}
 All sets $\parm(f)$ have the same size $n$.
\end{frameit}
\noindent
We also reserve the identifier $m$ for $|F|$.
  
\bdfn
The set of (abstract) program states is
$${\mathit St} = \{ (f,\sigma) \mid f\in F,\  \sigma:\parm(f)\to \val\}\,.$$
\edfn

Thus, a state is defined by a program point and a
\emph{store} $\sigma$ applicable to that point.  A state will be
cutomarily denoted by $s$ and we sometimes implicitly assume that its
components are $(f,\sigma)$.

A remark about the notion of abstract state may be in order.
While in simple settings (such as~\cite{leejonesbenamram01}),
$\parm(f)$ may correspond precisely to constituent variables of the
concrete program state, this is not true in general.
In many applications of
SCT, static program analysis is used to determine properties
of a state that are (or may be)
relevant to termination, e.g., the difference of two
integer variables, the depth of a recursion or closure stack, etc.
The original presentation of SCT referred to
programs that process data
types, such as lists and trees, that can be ranked by
their size, height etc; in this case we may prefer to think of the abstract
value
as a member of $\parm(f)$ rather than the list or tree itself. There are also examples where
it is worthwhile to include two abstractions (norms) of the same concrete
object.

\bdfn
For $f,g\in F$,
a size-change graph (SCG) with source $f$ and target $g$
is a bipartite directed graph with source nodes corresponding to $\parm(f)$
and target nodes corresponding to $\parm(g)$. We write this fact as $G:f\to g$.
Each arc of $G$ (called a \emph{size-change arc})
 is labeled with an element of the set
$\{\downarrow,\deq\}$.
\edfn

Size-change arcs represent constraints on \emph{state transitions}
$(f,\sigma)\mapsto (g,\sigma')$.
The arcs have the following meaning:

\begin{tabular}{rlll}
A \emph{strict} arc & $x\stackrel{\downarrow}{\to}y$ & represents the assertion
& $\sigma(x) > \sigma'(y)$ \\
A \emph{non-strict} arc & $x\stackrel{\deqsm}{\to} y$ & represents the assertion
& $\sigma(x) \ge \sigma'(y)$.
\end{tabular}
\medskip

We write $G\models (f,\sigma)\mapsto (g,\sigma')$ if
all constraints are satisfied; we say that transition $(f,\sigma)\mapsto
 (g,\sigma')$ is \emph{described by $G$}.

We write $x\to y\in G$ if there is an arc from $x$ to $y$ 
in $G$ (however labeled).

\medskip
An SCT instance, also known as 
\emph{annotated control-flow graph} (ACG),
is a CFG where every arc $c\in C_{fg}$
is annotated with a size-change graph $G_c:f\to g$. 

An SCT instance is customarily denoted by ${\mathcal G}$ and often viewed as
a set of SCG's, the CFG being implicitly specified.

\subsection{The SCT condition} \label{sec:SCT}

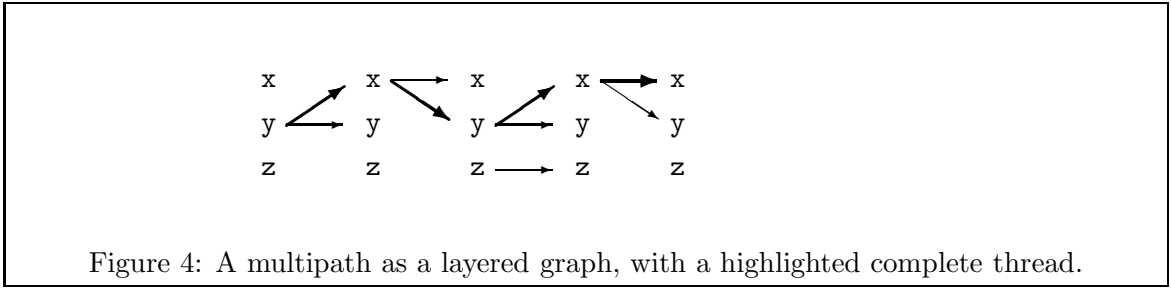
\begin{fig0}{A multipath as a layered graph, with a highlighted complete thread.}{t}{fig-multipath}
{\setlength{\unitlength}{0.4pt}
\flushleft\begin{picture}(600,200)(0,60)
\thinlines
	      \put(197,189){\tt x}
	      \put(197,147){\tt y}
	      \put(197,105){\tt z}
	      \thicklines
	      \put(221,153){\vector(3,2){55}}
	      \put(221,152){\vector(3,2){55}}
	      \thinlines
	      \put(221,153){\vector(1,0){53}}

	      \put(296,189){\tt x}
              \put(296,147){\tt y}
              \put(296,105){\tt z}
	      \thicklines
	      \put(320,195){\vector(3,-2){55}}
	      \put(320,194){\vector(3,-2){55}}
	      \thinlines
          \put(320,195){\vector(1,0){53}}

	      \put(395,189){\tt x}
              \put(395,147){\tt y}
              \put(395,105){\tt z}
	      \thicklines
	      \put(419,153){\vector(3,2){55}}
	      \put(419,152){\vector(3,2){55}}
	      \thinlines
 	      \put(419,110){\vector(1,0){53}}
	      \put(419,153){\vector(1,0){53}}
             
	      \put(494,189){\tt x}
              \put(494,147){\tt y}
              \put(494,105){\tt z}
	      \thicklines
              \put(518,195){\vector(1,0){53}}
	      \put(518,194){\vector(1,0){53}}
	      \thinlines
	      \put(518,195){\vector(3,-2){55}}
	      \put(584,189){\tt x}
              \put(584,147){\tt y}
              \put(584,105){\tt z}
\end{picture}} 
\end{fig0}

\bdfn
A $\mathcal G$-{\em multipath} is a sequence $M=G_1G_2\ldots$
of elements of ${\mathcal G}$ that label a (finite or
infinite) directed path in the CFG.
\edfn

The path in the CFG is
often denoted by $cs$ (the abbreviation stands for call sequence---originating
in a functional programming setting where transitions model calls).
  The multipath corresponding to $cs$ is
called $M_{cs}$. 
The \emph{front} of the
multipath is the source point of $G_1$, the \emph{rear} of a finite
multipath is the target of the last transition, and if both are one
and the same
flow-point, the multipath is referred to as \emph{a cycle}
(in fact, its underlying CFG path is a cycle).

We extend the notation $G\models s\mapsto s'$ to finite multipaths as follows:
\[
  G_1G_2\dots G_k \models s_0\mapsto s_k 
  \iff
  (\exists s_1,\dots,s_{k-1} )(\forall i) G_i \models s_i\mapsto s_{i+1}
\]

A multipath is ofen viewed as the (finite or infinite) {\em
layered directed graph\/}
obtained by identifying the target nodes of $G_i$ with the source
nodes of $G_{i+1}$ (Figure~\ref{fig-multipath}).

\bdfn
Let $M$ be a $\mathcal G$-multipath.
A {\em thread} in $M$ is a (finite or infinite) directed path in the
layered directed graph representation of $M$. We say that the thread is
\emph{from $x$ to $y$} if the thread begins with variable $x$ and ends with
$y$.

A thread is \emph{complete} if it spans the length of $M$.

A thread is {\em descending} if it includes a strict arc;
it is {\em infinitely descending} if it includes infinitely many strict arcs. 
\edfn

Intuitively, threads carry information along the computation,
 and this intuitive meaning
enters the language we are using, so for example we might say that 
variable $x$ is carried by a thread to variable $y$ (which implies that
the initial value of $x$ constrains the final value of $y$), or that a certain set
of variables is carried by threads to some other set.

More precisely, a thread represents a sequence of values generated during
a computation, that form a weakly decreasing chain in $\val$;
and an infinitely descending
thread indicates an infinitely-decreasing chain of values. Under the
assumption of well-foundedness of $\val$, such an infinite chain cannot
exist. This consideration leads to the following definition:

\bdfn[The SCT Condition]
$\mathcal G$ is said to \emph{satisfy SCT}, or \emph{be a positive SCT instance},
or  \emph{terminate},
if every infinite multipath contains
an infinitely-descending thread.
\edfn

We next formalize the manner in which the SCT condition (which is
purely combinatorial) relates to a semantic notion of termination.

\bdfn[$T_{\mathcal G}$]\label{def-transitionsystem}
The \emph{transition system} associated with ${\mathcal G}$
is the relation $T_{\mathcal G}$ over
$\mathit St$ defined by $$(s,s')\in T_{\mathcal G} \,\iff\, G\models s\mapsto s'
\mbox{ for some $G\in {\mathcal G}$}.$$
\edfn

We say that $T_{\mathcal G}$ is \emph{terminating} if there is no infinite chain in $T_{\mathcal G}$.

\begin{thm}%[SCT main theorem]
 \label{thm-sct}
$T_{\mathcal G}$ is terminating if and only if ${\mathcal G}$ satisfies SCT.
\end{thm}

The ``if'' part of this theorem (soundness of the SCT criterion) follows
directly from well-foundedness.
For the ``only if'' direction
see~\cite{Lee:ranking}.
Thus, SCT is a sound and complete termination criterion for the corresponding
class of transition system.

SCT is decidable (it would be a far less
interesting abstraction otherwise); note that this is possible because
the corresponding transition systems are restricted and only approximate real
programs.
A well-known way to decide SCT is the so-called
\emph{closure algorithm}, which consists of computing a transitive closure
of ${\mathcal G}$ and checking idempotent graphs: we now review the pertinent
definitions and facts, following~\cite{leejonesbenamram01}.

\bdfn \label{def-composition}
The {\em composition} of size-change graph $G_1:{\tt
f}\to{\tt g}$ with $G_2:{\tt g}\to{\tt h}$ is a size-change graph with source
{\tt f}, target {\tt h} and
arc set $E^\downarrow\cup E\stackrel{\!\deqsm}{}$, where

$$\begin{array}{lll}        
	E^\downarrow &=& \{x\stackrel{\downarrow}{\to}z\sep 
		x\stackrel{r_1}{\to}y\in G_1,\ y\stackrel{r_2}{\to}z\in G_2,\
		r_1\mbox{ or }r_2\mbox{ is}\downarrow\}\\
    E\stackrel{\!\deqsm}{} &=& \{x\stackrel{\deqsm}{\to}z\sep 
		x\stackrel{\deqsm}{\to}y\in G_1,\ 
		y\stackrel{\deqsm}{\to}z\in G_2,\
		x\stackrel{\downarrow}{\to}z\notin E^\downarrow\}.
\end{array}$$

The composition is denoted by $G_1;G_2$.
\edfn

\bdfn \label{def-idempotent}
Graph $G$ is \emph{idempotent} if $G;G=G$.
\edfn

Note that an idempotent graph must have the same flow-point for both
source and target, i.e., it describes a cycle in the control-flow graph.

\bthm
Let ${\mathcal G}^+$ denote the composition-closure of ${\mathcal G}$.
SCT is satisfied by ${\mathcal G}$ if and only if every idempotent graph in
${\mathcal G}^+$ has an arc $x\stackrel{\downarrow}{\to} x$ for some $x$.
\ethm

A variation in which every graph in the closure
(regardless of idempotence) is tested
is described in~\cite{Sa:91,Codish-et-al:05}.

\subsection{Ranking functions}

\bdfn
Let $\mathcal T$ be a transition system over state-space {\it St}.
A (global) \emph{ranking function} for $\mathcal T$
is a function $\rho:{\mathit St}\to W$, where $W$ is
a well-founded set,
such that $(s,s')\in {\mathcal T}\Rightarrow \rho(s) > \rho(s')$.
\edfn

\noindent
Let $P(s,s')$ be any predicate, where $s,s'$ are free variables representing states. 
We write $G\models P(s,s')$ if
$$\forall s,s' \quad (G\models s\mapsto s') \Rightarrow P(s,s').$$

\bdfn
A (global) \emph{ranking function} for an SCT instance $\mathcal G$
is a function $\rho:{\mathit St}\to W$, where $W$ is
a well-founded set,
that constitutes a ranking function for $T_{\mathcal G}$. Equivalently, it
satisfies $G\models \rho(s) > \rho(s')$ for every $G\in {\mathcal G}$.
\edfn

\noindent
For convenience, we often ``Curry'' $\rho$ and
write $\rho(f,\sigma)$ as $\rho_f(\sigma)$.

\paragraph*{Complexity measures.} In this paper we are interested in
explicit construction of ranking functions. Thus the ranking functions will
be given by expressions, combining the values of program variables with
appropriate operators (such as {\bf min}, {\bf max} etc.). The complexity
measure we are mostly interested in is the size of the expression.

Also of interest is the time to construct the expression (if the proof
is constructive).
Naturally, this time is lower-bounded by the expression's size.

\paragraph*{Notations.} If a flow point $f$
has variables $x,y,\dots$, then it is
natural to write a ranking function using these variable names,
e.g., $\rho_f(x,y,\dots) = \max(x,y,\dots)$. But technically, a ranking function is a
function over ${\mathit St}$. To iron out the formality, we use notations as
defined next.

\bdfn
Let $S\subseteq \parm(f)$. Then $\maxval{\sigma}{S}$ is
$\max\{\sigma({\tt x})\mid {\tt x}\in S\}$. For fixed $S$, this is a function of $\sigma$.
Similarly, $\minval{\sigma}{S}$ is $\min\{\sigma({\tt x})\mid {\tt x}\in S\}$.
\edfn

\subsection{Subclasses of SCT}

Previous work on SCT has identified certain structural subclasses
of SCT as interesting. By structural, we mean that the subclass is defined
by imposing a constraint on the structure of the size-change graphs.
The following three subclasses will play a role in this work:

\begin{enumerate}[$\bullet$]
\item
In \emph{Fan-in free} graphs, the in-degree of all nodes is at most 1.
Fan-in free graphs are discussed in~\cite{BL:2006,BA:delta}.
A benchmark evaluation described in~\cite{BL:2006} demonstrated that such
graphs occur frequently when size-change graphs are extracted from 
Prolog programs. In \cite{BA:delta}, it was shown
that fan-in freedom makes an extended (and harder) form of SCT,
 called $\delta$SCT, decidable.
\item
In \emph{Fan-out free} graphs, the out-degree of all nodes is at most 1.
The interest in this subclass is mostly due to its being defined	
symmetrically to fan-in free graphs, and yet sometimes easier to work with.
For example, in this paper we will handle fan-out free graphs first.
\item
\emph{Strict SCT} graphs have exclusively strict arcs. This again is a class
which is introduced because it is easier to work with. In particular,
as shown
in~\cite{Lee:ranking}, this class admits a simple ranking-function
construction; we will make use of that construction in this work.
\end{enumerate}

\subsection{Multiset orderings} \label{sec-ms}

In the example given in Section~1, as well as in many classical examples,
it is useful to define the rank of a state as a tuple that is shown to descend lexicographically.
In previous work on termination, it has been discovered that in some cases,
it is useful to form not a tuple, but a multiset of values, and exhibit descent in an appropriate
\emph{multiset order}. In particular, such orders turn out useful in constructing ranking
functions for SCT; this subsection presents the necessary definitions.

\bdfn[Simple Multiset Order, SMO]
Let $A,B$ be finite multisets over $\val$.
We write $A>B$ if $|A|>|B|$ (the cardinality of $A$ is larger) or
if $|A|=|B|$ and the sets can be listed as $A=\{a_1, a_2,\dots\}$
and $B=\{b_1,b_2,\dots\}$ with $a_i\ge b_i$ for all $i$ and $a_i>b_i$ for
at least one $i$. We write $A\ge B$ for the non-strict variant.
\edfn

For example, 
let $A = \{4,3,3,0\}$ and $B = \{4,3,2,0\}$. Then we have $A > B$.
If we let $C = \{4,3,2,1\}$, we have neither $A\ge C$, nor $C\ge A$: these
sets are incomparable. SMO is thus a partial order.

Let $k>0$ and suppose that $|A|=|B|=k$; it is easy to verify that 
 $A > B$ means that a sorted listing of $A$ is
lexicographically greater than a sorted listing of $B$. This is true for both
descending sort and ascending sort; which suggests two ways of completing SMO
to a total order over $k$-element multisets.
To complete the last example, since $\langle 4,3,2,1\rangle$ is lexicographically
smaller than $\langle 4,3,3,0\rangle$, multiset $C$ is smaller according to descending sort.
However, according to ascending sort, we find that $C$ is greater, as
 $\langle 1,2,3,4\rangle$ is lexicographically greater than $\langle 0,3,3,4\rangle$.

Comparing lists in descending sorted order yields the total order called
\emph{multiset order} by Dershowitz and Manna, who defined it in a more general
fashion and showed its usefulness in termination proofs. Comparing ascending lists
results in the so-called \emph{dual multiset order}~\cite{BL:2006}. 
In this work, we will make use of both total orders.
We emphasize, however, that we only use them for sets of equal
cardinality (which simplifies their definitions).
 If $|A|>|B|$, we shall always consider $A$ to be bigger.

A total order is needed, in particular, for the definition of
min and max operators.

\bdfn \label{def-minmax}
Let $A,B$ be finite multisets over $\val$. We
define $\min(A,B)$ and $\max(A,B)$ as follows: 
if $|A|\ne |B|$, then $\min(A,B)$ is the
smaller multiset and $\max(A,B)$ is the larger.
 If $|A|=|B|$, $\min(A,B)$ is the smaller under
dual multiset order, while $\max(A,B)$ is the bigger under 
multiset order.
\edfn

The definitions extend naturally to define minimum and maximum over a finite
set of multisets. In all cases, the operator can be implemented by choosing
the lexicographic minimum, or maximum, among the tuples that represent the
multisets (with elements in ascending order for {\bf min}, descending order
for {\bf max}).
The choice of two different orderings to define {\bf min} and
{\bf max} may seem strange, but it will be seen to work best later in the paper.

\section{Some Previous Results}

This section summarizes some previous work on SCT,
including definitions and constructions that we shall use.

\subsection{Thread preservers}

The concept of thread preservers was introduced in~\cite{BL:2006}.
We cite the definition and a useful theorem.

Let ${\mathcal V} = \bigcup_{f}\parm(f)$,
the combined set of variables in 
the whole abstract program.

\bdfn%[Thread preserver]
Let $\mathcal G$ be an ACG and $\mathcal V$ its set of variables.
A set $P\subseteq{\mathcal V}$ is called a {\em thread preserver\/} for $\mathcal G$
if for every $G\in {\mathcal G}$ where $G:f\to g$, it holds that
whenever $x\in (\parm(f)\cap P)$,
there is $x{\to}y\in G$  for some $y\in P$.
\edfn

It is easy to see that the set of thread preservers of a given ACG
is closed under
union. Hence, there is always a unique maximal thread preserver (MTP) for 
$\mathcal G$, which we denote by $\mbox{MTP}({\mathcal G})$.
It is further shown in \cite{BL:2006} that
given a standard representation of $\mathcal G$, 
$\mbox{MTP}({\mathcal G})$ can be found in linear time.
This is significant because the MTP is useful---among else for constructing
ranking functions.

\subsection{Ranking functions for strict SCT}
\label{sec-strict}

Throughout this subsection, ${\mathcal G}$ is presumed strongly connected.

The following theorem is from~\cite{Lee:ranking}:

\bthm \label{thm-subsets}
Let $\mathcal G$ be a terminating, strict SCT instance.
There exists an indexed set $\{S_f\}$, where for every $f\in F$, $S_f$
is a set of subsets of $\parm(f)$, such that the function
$$\rho(f,\sigma) = \min_{X\in S_f} \maxval{\sigma}{X}$$
is a ranking function for $\mathcal G$.
\ethm

For example, consider the SCT instance in Figure~\ref{fig-ex-m05}.
It has the ranking function 
$$\rho(f,\sigma) = \min( \maxval{\sigma}{\{{\tt x},{\tt y}\}}, 
                                       \maxval{\sigma}{\{{\tt x},{\tt z}\}},
                                       \maxval{\sigma}{\{{\tt y},{\tt z}\}} ) .$$
The correctness of this function can be verified, as usual, by checking graph by graph
and assuming in turn each ordering of the values of the variables.

\begin{fig0}{An SCT instance with strict arcs only}{t}
{fig-ex-m05}
{\setlength{\unitlength}{0.46pt}\footnotesize
\begin{center}
\begin{picture}(340,200)(120,-15)
	      \put( 85,20){\framebox(120,125){}}
              \put( 95,110){${\tt x}$}
              \put( 95, 70){${\tt y}$}
              \put( 95, 30){${\tt z}$}
	      \thicklines
	      \put(115,73){\vector(1,0){50}}
	      \put(115,72){\vector(1,0){50}}
	      
	      \put(115,33){\vector(1,0){50}}
	      \put(115,32){\vector(1,0){50}}
	      
	      \put(115,73){\vector(4,-3){50}}
	      \put(115,72){\vector(4,-3){50}}
	      
	      \put(115,33){\vector(4,3){50}}
	      \put(115,32){\vector(4,3){50}}
	      \thinlines
              \put(175,110){${\tt x}$}
              \put(175, 70){${\tt y}$}
              \put(175, 30){${\tt z}$}

	      \put(240,20){\framebox(120,125){}}
              \put(250,110){${\tt x}$}
              \put(250, 70){${\tt y}$}
              \put(250, 30){${\tt z}$}
	      \thicklines
	      \put(270,115){\vector(1,0){50}}
	      \put(270,114){\vector(1,0){50}}
	      
	      \put(270,115){\vector(2,-3){50}}
	      \put(270,114){\vector(2,-3){50}}
	      
	      \put(270,33){\vector(1,0){50}}
	      \put(270,32){\vector(1,0){50}}
	      
	      \put(270,33){\vector(2,3){50}}
	      \put(270,32){\vector(2,3){50}}
	      \thinlines
              \put(335,110){${\tt x}$}
              \put(335, 70){${\tt y}$}
              \put(335, 30){${\tt z}$}

	      \put(395,20){\framebox(120,125){}}
              \put(405,110){${\tt x}$}
              \put(405, 70){${\tt y}$}
              \put(405, 30){${\tt z}$}
	      \thicklines
	      \put(425,115){\vector(1,0){50}}
	      \put(425,114){\vector(1,0){50}}
	      
	      \put(425,115){\vector(4,-3){50}}
	      \put(425,114){\vector(4,-3){50}}
	      
	      \put(425,73){\vector(1,0){50}}
	      \put(425,72){\vector(1,0){50}}

	      \put(425,73){\vector(4,3){50}}
	      \put(425,72){\vector(4,3){50}}
	      \thinlines
              \put(490,110){${\tt x}$}
              \put(490, 70){${\tt y}$}
              \put(490, 30){${\tt z}$}

	      \put(100,-5){$G_1: f\to f$}
	      \put(255,-5){$G_2: f\to f$}
	      \put(410,-5){$G_3: f\to f$}
\end{picture}\end{center}\par}
%\vair
\end{fig0}

%%%%%%%%%%%%%

The worst-case size of a function of the above form
is exponential, related to the number of different
subsets of $\parm(f)$, which is $2^n$.
For more details, see~\cite[\S~3.1]{Lee:ranking}.  Next, we give two special cases
of particular interest,
which are already implicit in~\cite{BL:2006}. The first case
is that of \emph{fan-out free} graphs:

\bthm \label{thm-min}
If $\mathcal G$ is a terminating, strict, fan-out free SCT instance,
then $\mathcal G$ has a non-empty thread-preserver; and for any such preserver $P$,
the function
$$\rho(f,\sigma) = \minval{\sigma}{(\parm(f)\cap P)} $$
is a ranking function for $\mathcal G$.

\ethm

The theorem follows from the following two lemmas. The first we only cite,
referring the reader to~\cite[\S~6]{BL:2006} for a proof.  The second one we
prove, because the proof is simple and clarifies the significance of thread
preservers in connection with ranking.

\blem \label{lem-fofTP}
Suppose that $\mathcal G$ is strict and fan-out free.
Then $\mathcal G$ is size-change terminating if and only if ${\mathcal G}$ has a non-empty thread-preserver. 
\elem

\blem
Suppose that $\mathcal G$  has a non-empty thread preserver $P$,
and let
$$\rho(f,\sigma) = \minval{\sigma}{(\parm(f)\cap P)} \,. $$
For all $G\in {\mathcal G}$, $G\models \rho(s)\ge \rho(s')$.
If $\mathcal G$ is strict, then $G\models \rho(s) > \rho(s')$.
\elem

\bprf
Let $G\models (f,\sigma)\mapsto (f',\sigma')$.
By the definition of a thread preserver, we have, for all 
 $x\in (\parm(f)\cap P)$,
$\sigma(x)\ge \sigma'(y)$  for some $y\in \parm(f')\cap P$.
If $x$ is such that $\sigma(x)$ is minimum
(i.e., $\minval{\sigma}{(\parm(f)\cap P)}$ is $\sigma(x)$),
we have $\rho(f,\sigma)=\sigma(x)$, whereas
$\rho(f',\sigma')\le\sigma(y)$.  Combining the three relations,
we get $\rho(f',\sigma')\le\rho(f,\sigma)$.

If $\mathcal G$ is strict, then $\sigma(x) > \sigma'(y)$ and
the inequality becomes strict.
\eprf

The case of \emph{fan-in free graphs} is symmetric to that of fan-out free
graphs. In order to exploit this symmetry, we use the technique of 
\emph{transposing} the size-change graphs.

\bdfn
If $G:f\to g$ is a size-change graph, $G^t$ denotes its transposition,
which is a size-change graph with source $g$, target $f$, and arcs
$\{ y\to x \mid x\to y \in G \}$.
For a set $\mathcal G$ of size-change graphs, 
	${\mathcal G}^t = \{G^t\mid G\in{\mathcal G}\}$.
\edfn

\begin{obs}[\cite{BL:2006}]\label{obs-transp}
$\mathcal G$ satisfies SCT if and only if ${\mathcal G}^t$ does.
\end{obs}

Clearly, $G$ is fan-in free if and only if $G^t$ is fan-out free.
Now we can use this for a ranking-function construction.

\bthm \label{thm-max}
If $\mathcal G$ is a terminating, strict, fan-in free SCT instance,
then ${\mathcal G}^t$ has a non-empty thread-preserver; and for any such preserver $P$,
the function
$$\rho(f,\sigma) = \maxval{\sigma}{(\parm(f)\cap P)} $$
is a ranking function for $\mathcal G$.
\ethm

The fact that ${\mathcal G}^t$ has a non-empty thread-preserver follows from
Observation~\ref{obs-transp} and Lemma~\ref{lem-fofTP}.
The correctness of the ranking function then follows from the next lemma.

\blem
Suppose that ${\mathcal G}^t$ has a non-empty thread preserver $P$.
And let
$$\rho(f,\sigma) = \maxval{\sigma}{(\parm(f)\cap P)} \,. $$
For all $G\in {\mathcal G}$, $G\models \rho(s)\ge \rho(s')$.
If $\mathcal G$ is strict, then $G\models \rho(s) > \rho(s')$.
\elem

\bprf
Let $G\models (f,\sigma)\mapsto (f',\sigma')$.
By the definition of a thread preserver (but noting that it is
a $G^t$ thread preserver!), we have, for all 
 $y\in (\parm(f')\cap P)$,
$\sigma(x)\ge \sigma'(y)$  for some $x\in \parm(f)\cap P$.
For $y$ such that $\sigma(y)$ is maximum,
we have $\rho(f',\sigma')=\sigma(y)$, whereas
$\rho(f,\sigma)\ge\sigma(x)$.  Combining the three relations, 
we get $\rho(f,\sigma)\ge\rho(f',\sigma')$.

If $\mathcal G$ is strict, then $\sigma(x) > \sigma'(y)$ and
the inequality becomes strict.
\eprf

\section{Fan-out Free SCT} \label{sec-fof}

We are given a fan-out free positive
SCT instance ${\mathcal G}$.
We assume that ${\mathcal G}$ is strongly connected
and that $|\parm(f)|=n$ for all $f\in F$, and let $m = |F|$.
We shall construct a ranking function for $\mathcal G$, and then discuss its size
(getting an upper bound which is later shown to be tight).
A brief outline of the construction follows:
\begin{enumerate}[$\bullet$]
\item
Transform $\mathcal G$ into a strict, fan-out free SCT instance.
\item
Use Theorem~\ref{thm-min}.
\item
Optionally, optimize the ranking function for size (without this stage, the desired upper
bound on size may fail to hold).
\end{enumerate}

\noindent Next, the construction is described in detail, along with
proofs and some illustrations.  A complete demonstration of the
process for a (very small) example can be found at the end of the
section, along with some comments regarding the implementation of the
algorithm (which is initially described in a very abstract manner,
just to make the theorems clear).

\subsection{The basic construction} \label{sec:basic-fof}

\bdfn[Vectors] \label{def-Vf}
For flow-point $f\in F$ and positive integer $B$, 
 $V_f^B$ is the set of tuples 
$\vec v_f = \langle v_1,v_2,\dots\rangle$ of even length,
where every odd position is a non-empty subset of 
$\parm({f})$, together
constituting a partition of the latter;
and every even position is an integer between 0 and $B$.
\edfn

Remark. The distinction between a vector and its components is indicated here by font.
Thus, $v_i$ is the $i$th entry of $\vec v$, while $\vec v_i$ is the $i$th vector in some sequence of vectors.
For notational
convenience, we may make use of a double-meaning expression $|v_i|>|u_i|$,
which means, if $i$ is odd, that the set $v_i$ contains more elements than
$u_i$; and if $i$ is even, that the integer $v_i$ is greater.

\bdfn \label{def-value}
The value of $\vec v \in V_f^B$ in a given program state $(f,\sigma)$,
denoted $\sigma(\vec v)$, is obtained by
substituting the values of variables according to $\sigma$,
so every subset of variables becomes a multiset
of data values. This results in a tuple with multisets and integers in odd and even
positions, respectively. Such tuples are compared lexicographically,
where multisets are compared according to one of
the multiset orders (we use SMO, as long as a total order is not required).
We define the {\bf min} and {\bf max} operators
on vectors by the lexicographic extension of the corresponding total multiset
order, according to Definition~\ref{def-minmax}.
\edfn

In the rest of this section, $B=m\cdot 2^n$. Since $B$ is
fixed, $V_f^B$ may be written as $V_f$.

\bdfn
For $S\subseteq \parm(f)$ and a size-change
graph $G:f\to g$, define $\Im(S,G)$ to be the
set of $y$ such that $x\to y\in G$ with $x\in S$. 
\edfn
Observe that, as we are only dealing with fan-out free graphs in this
construction, $|\Im(S,G)|\le |S|$, with equality only if we have a 
\emph{one-to-one} correspondence, where for every element of $S$ (respectively
$\Im(S,G)$) there is a single arc in $G$ connecting it to the other set.
It is quite easy to see, that such correspondence implies a (weak)
SMO descent from $S$ to $\Im(S,G)$ (if at least one of
the connecting arcs is strict, we have strict descent).

\bdfn
Let $\vec v \in V_f$, and $G:f \to g$.
For an odd position $i$ in $\vec v$, let 
$$\Im_i(\vec{v},G) = \Im(v_i,G) \setminus \bigcup_{\textrm{odd\ }j<i} \Im(v_j,G).$$

We say that position $i$ is {\em descending} if $|\Im_i(\vec{v},G)|<|v_i|$ or
$|\Im_i(\vec{v},G)|=|v_i|$ and there is a strict arc in $G$ from $x\in v_i$ to
$y\in\Im_i(\vec{v},G)$. 
\edfn
Note that the only other possibility is $|\Im_i(\vec{v},G)|=|v_i|$ 
and all arcs from $v_i$ to $\Im_i(\vec{v},G)$ non-strict. So, a descending position's value
either strictly descends in multiset order, or at least weakly descends.
In the next definition,
we use the numeric positions in order to make the vector decrease strictly when there is no
descent in the set-valued positions. We also use them to avoid having empty sets
in the set-valued positions.

\bdfn \label{def-next}
Given a size-change graph $G:f\to g$, and $\vec v\in V_f$,
$\nxt(\vec v, G)\in V_g$ is defined by cases,
as follows. 
\begin{enumerate}[\enspace]
\item\noindent{\bf Case N1:}\ If $i$ is the first
descending position and $|\Im_i(\vec{v},G)| > 0$,
$$\nxt(\vec v,G) = 
	\langle \Im_1(\vec{v},G), v_2, \Im_3(\vec{v},G), \dots, 
\Im_i(\vec{v},G), B, S, B\rangle$$ where 
$S$ is the set of $g$ variables not occurring up to position $i$;
if $S$ turns out to be empty, omit the final suffix $S,B$.\medskip

\item\noindent{\bf Case N2:}\
If $i$ is the first descending position and $|\Im_i(\vec{v},G)| = 0$,
$$\nxt(\vec v, G) = 
	\langle \Im_1(\vec{v},G), v_2, \Im_3(\vec{v},G), \dots, 
	\Im_{i-2}(\vec{v},G), v_{i-1}-1, S, B\rangle
$$ 
where $S$ is the set of $g$ variables not occurring up to position
$i-2$; note that it cannot be empty.  The above expression assumes
that $i>1$ and $v_{i-1} > 0$. If either of these conditions is not
met, $\nxt(\vec v, G)$ is \emph{undefined}.\medskip

\item\noindent{\bf Case N3:}\
No position is descending. The last position is $v_i$ (an integer).
$$\nxt(\vec v, G) = 
	\langle \Im_1(\vec{v},G), v_2, \Im_3(\vec{v},G), \dots, 
	\Im_{i-1}(\vec{v},G), v_{i}-1\rangle
$$
assuming that $v_i>0$; if $v_i=0$, $\nxt(\vec v, G)$ is
\emph{undefined}.\medskip
\end{enumerate}
\edfn

Note that a size-change arc of $G$ that leaves a variable
$x\in v_i$ never
reaches a variable in a higher position of $\nxt(\vec v, G)$; it may
reach position $i$ or a lower one. Here is a more substantial observation:

\begin{obs} \label{obs-next}
In a program transition $(f,\sigma)\stackrel{c}{\to}(g,\sigma')$,
the value of $\nxt(\vec v, G_c)$ in $(g,\sigma')$ (if defined)
is strictly smaller than the value of $\vec v$ in $(f,\sigma)$.
\end{obs}

Thus, we can use these vectors to construct an instance of strict
SCT.
Note that the descent is justified by the definition of next and
by size-change graph $G_c$. Therefore, a ranking function built
on the base of this descent can be statically verified to decrease,
based on ${\mathcal G}$.

\bdfn \label{def:R}
$\mathcal R$ is an all-strict, fan-out free
SCT instance with CFG as in $\mathcal G$,
where the variables for flow-point $f$ are
the vectors $V_f$, and the size-change graph $G_c^R$ for 
arc $c\in C_{fg}$ has arcs $\vec v \to \nxt(\vec v, G_c)$ for all $\vec v\in V_f$
such that $\nxt(\vec v, G_c)$ exists.
\edfn

Clearly, $\mathcal R$ is fan-out free. Note that this implies that, given a multipath
and a specific vector at its source, a unique thread can be followed from
that vector until it either stops at the end of the multipath or reaches
a vector with no outcoming arc.
Such a thread is a chain of vectors obtained by repeated application of $\nxt$.
For an example, 
Figure~\ref{fig-ex-Rgraphs} shows three size-change graphs of an instance $\mathcal G$
and Figure~\ref{fig-ex-Rthread}
shows a single thread from an $\mathcal R$-multipath.
Note that there is lexicographic descent at each step.

\def\doublestack#1#2#3{\mathrel{\mathop{#2}\limits^{#1}_{#3}}}

\begin{fig1}{A fan-out free SCT instance.}
{t}{fig-ex-Rgraphs}
{\setlength{\unitlength}{0.4pt}
\flushleft\begin{picture}(600,200)(0,60)
\thinlines
	      \put(188,85){\framebox(120,138){}}
	      \put(240,60){$G_1$}
	      \put(197,189){\tt x}
	      \put(197,147){\tt y}
	      \put(197,105){\tt z}
	      \thicklines
	      \put(221,195){\vector(3,-2){55}}
	      \put(221,194){\vector(3,-2){55}}
	      \put(221,153){\vector(1,0){53}}
	      \put(221,152){\vector(1,0){53}}
	      \thinlines
	      \put(287,189){\tt x}
	      \put(287,147){\tt y}
	      \put(287,105){\tt z}
	      
	      \put(350,85){\framebox(120,138){}}
	      \put(402,60){$G_2$}
	      \put(359,189){\tt x}
              \put(359,147){\tt y}
              \put(359,105){\tt z}
	      \thinlines
          \put(383,195){\vector(1,0){53}}
	      \put(383,153){\vector(3,2){55}}
	      \put(383,111){\vector(1,0){53}}
	      \put(449,189){\tt x}
              \put(449,147){\tt y}
              \put(449,105){\tt z}

	      \put(512,85){\framebox(120,138){}}
	      \put(564,60){$G_3$}
	      \put(521,189){\tt x}
              \put(521,147){\tt y}
              \put(521,105){\tt z}
	      \thicklines
              \put(545,111){\vector(1,0){53}}
	      \put(545,110){\vector(1,0){53}}
	      \put(545,195){\vector(3,-2){55}}
	      \put(545,194){\vector(3,-2){55}}
	      \thinlines
	      \put(545,153){\vector(1,0){53}}
	      \put(611,189){\tt x}
              \put(611,147){\tt y}
              \put(611,105){\tt z}
\end{picture}} 
\end{fig1}

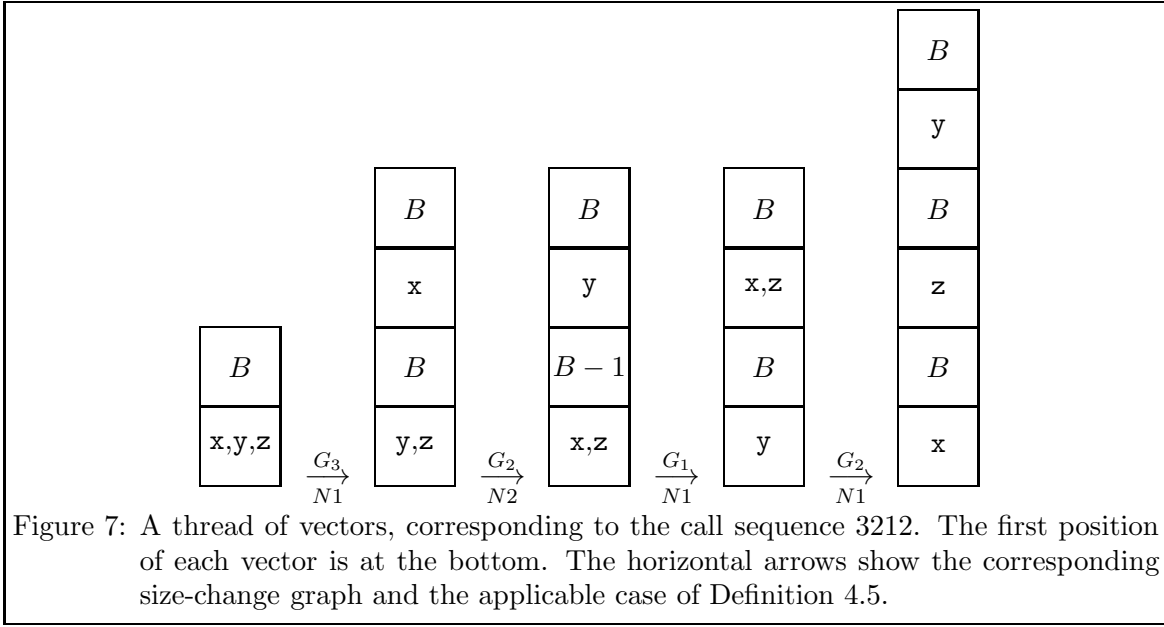
\begin{fig0}{A thread of vectors, corresponding to the 
call sequence 3212. The first position of each vector is at the bottom.
The horizontal arrows show the
corresponding size-change graph and the applicable case of Definition~4.5.}
{t}{fig-ex-Rthread}
{\setlength{\unitlength}{0.3pt}
\begin{center}\begin{picture}(1000,600)(0,0)
\thinlines
	      \put(0,  0){\framebox(100,100){{\tt x},{\tt y},{\tt z}}}
	      \put(0,100){\framebox(100,100){$B$}}
	
		  \put(100, -20){\makebox(120,60){\makebox[36pt][c]{$\doublestack{G_3}{\longrightarrow}{N1}$}}}
      
	      \put(220,  0){\framebox(100,100){{\tt y},{\tt z}}}
	      \put(220,100){\framebox(100,100){$B$}}
	      \put(220,200){\framebox(100,100){{\tt x}}}
	      \put(220,300){\framebox(100,100){$B$}}
	      
		  \put(320,-20){\makebox(120,60){\makebox[36pt][c]{$\doublestack{G_2}{\longrightarrow}{N2}$}}}
      
	      \put(440,  0){\framebox(100,100){{\tt x},{\tt z}}}
	      \put(440,100){\framebox(100,100){$B-1$}}
	      \put(440,200){\framebox(100,100){{\tt y}}}
	      \put(440,300){\framebox(100,100){$B$}}

		  \put(540,-20){\makebox(120,60){\makebox[36pt][c]{$\doublestack{G_1}{\longrightarrow}{N1}$}}} 
	      
	      \put(660,  0){\framebox(100,100){{\tt y}}}
	      \put(660,100){\framebox(100,100){$B$}}
	      \put(660,200){\framebox(100,100){{\tt x},{\tt z}}}
	      \put(660,300){\framebox(100,100){$B$}}
	      
		  \put(760,-20){\makebox(120,60){\makebox[36pt][c]{$\doublestack{G_2}{\longrightarrow}{N1}$}}}
      
	      \put(880,  0){\framebox(100,100){{\tt x}}}
	      \put(880,100){\framebox(100,100){$B$}}
	      \put(880,200){\framebox(100,100){{\tt z}}}
	      \put(880,300){\framebox(100,100){$B$}}
	      \put(880,400){\framebox(100,100){{\tt y}}}
	      \put(880,500){\framebox(100,100){$B$}}
\end{picture}\end{center}} 
\end{fig0}

%%%%%%%%%%%%%%%%%%%%

We now introduce some notation and concepts used later in analysing $\mathcal R$.
Let $M^G = G_1G_2\dots$ denote a $\mathcal G$-multipath, and
$M^R$ the corresponding $\mathcal R$-multipath (following the same CFG path).
Let $\tau$ denote a thread in $M^R$, namely
a sequence $\vec v_0,\vec v_1,\dots$
such that $\vec v_j = \nxt(\vec v_{j-1}, G_{j})$.
The passage from $\vec v_{j-1}$ to $\vec v_j$ will be called \emph{a
step} and written compactly as $\vec v_{j-1}\stackrel{G_{j}}{\to}\vec v_j$.

\bdfn
The $i$-band of $\tau$ is the sequence resulting from trimming all
vectors in $\tau$ to their $i$-prefix, i.e., the first $i$ positions.
We say that $\tau$ is $i$-stable if for all steps
$\vec v\stackrel{G}{\to}\vec v'$ in $\tau$, no odd position $k\le i$
is descending.
\edfn

We denote by $P_i(\vec v)$ the set of variables appearing in the $i$-prefix
of $\vec v$.

% ex-rf-fof.tex  continued

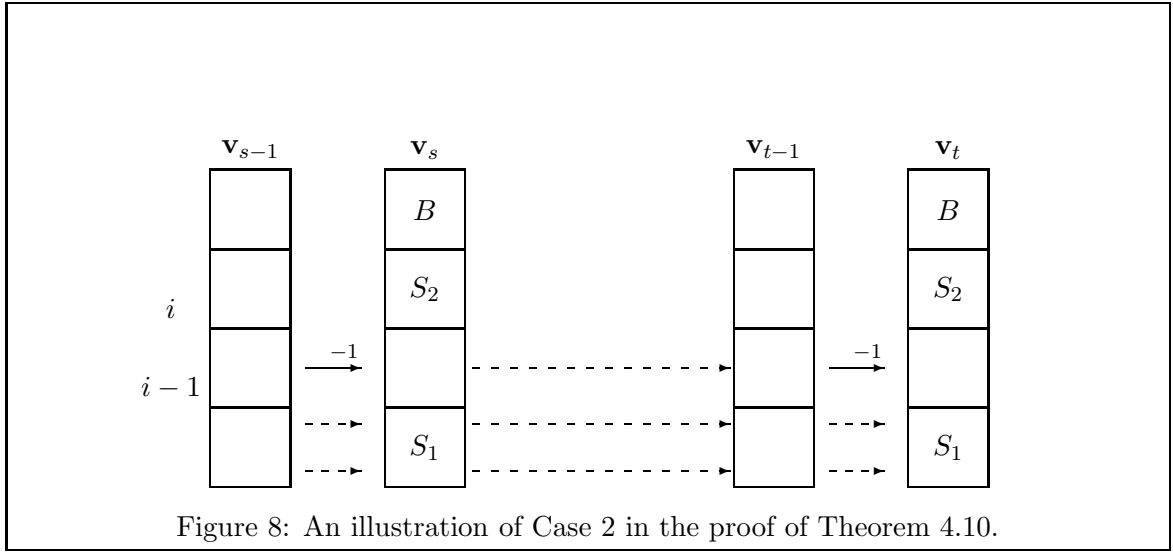
\begin{fig0}{An illustration of Case~2 in the proof of Theorem~4.10.}
{t}{fig-Case2}
{\setlength{\unitlength}{0.3pt}
\begin{center}\begin{picture}(1000,600)(-10,0)
\thinlines
		  \put(-100, 100){\makebox(100,50){$i-1$}}
		  \put(-100, 200){\makebox(100,50){$i$}}
	      \put(0,  0){\framebox(100,100){}}
	      \put(0,100){\framebox(100,100){}}
		  \put(0,200){\framebox(100,100){}}
		  \put(0,300){\framebox(100,100){}}
		  \put(0,400){\makebox(100,50){${\vec v}_{s-1}$}}

		  \dashline{10}(120, 20)(190,20)
		  \put(180,  20){\vector(1,0){10}}
		  \dashline{10}(120, 80)(190,80)
		  \put(180,  80){\vector(1,0){10}}
		  \put(120, 150){\vector(1,0){70}}
		  \put(150, 150){${}^{-1}$}
      
	      \put(220,  0){\framebox(100,100){$S_1$}}
	      \put(220,100){\framebox(100,100){}}
	      \put(220,200){\framebox(100,100){$S_2$}}
	      \put(220,300){\framebox(100,100){$B$}}
		  \put(220,400){\makebox(100,50){${\vec v}_{s}$}}
	      
		  \dashline{10}(330,20)(650,20)
		  \put(645,20){\vector(1,0){10}}
		  \dashline{10}(330,80)(650,80)
		  \put(645,80){\vector(1,0){10}}
		  \dashline{10}(330,150)(650,150)
		  \put(645,150){\vector(1,0){10}}
      
	      \put(660,  0){\framebox(100,100){}}
	      \put(660,100){\framebox(100,100){}}
	      \put(660,200){\framebox(100,100){}}
	      \put(660,300){\framebox(100,100){}}
		  \put(660,400){\makebox(100,50){${\vec v}_{t-1}$}}
	      
		  \dashline{10}(780, 20)(850,20)
		  \put(840,  20){\vector(1,0){10}}
		  \dashline{10}(780, 80)(850,80)
		  \put(840,  80){\vector(1,0){10}}
		  \put(780, 150){\vector(1,0){70}}
		  \put(810, 150){${}^{-1}$}

	      \put(880,  0){\framebox(100,100){$S_1$}}
	      \put(880,100){\framebox(100,100){}}
	      \put(880,200){\framebox(100,100){$S_2$}}
	      \put(880,300){\framebox(100,100){$B$}}
		  \put(880,400){\makebox(100,50){${\vec v}_{t}$}}
\end{picture}\end{center}} 
\end{fig0}
%%%%%%%%%%%%%%%%%%%%%%%%%%%%%%%%%%%%%%

\blem
Suppose that $i$ is even, and $\tau$ is an $\mathcal R$-thread with a stable
$i$-band. Then none of the numeric (even) positions below $i$ changes
along $\tau$, while the last one (position $i$) may decrease, but cannot
increase.
\elem

\bprf
Straight-forward from definition of the $\nxt$ relation
and $i$-stability.
\eprf

On the other hand, if the $i$-band is not stable in $\tau$, then either position
$i$ or a
lower numeric position must be reset to $B$ once or more along $\tau$.

\bthm \label{thm:R}
$\mathcal R$ satisfies SCT.
\ethm

\bprf
Let $f_0$ be the initial function of $\mathcal G$ and let
$\vec v_0 = \langle \parm({f_0}), B\rangle$.
We claim that in every $\mathcal R$ multipath (in particular, an infinite one)
starting at $f_0$ there is a complete thread 
starting at $\vec v_0$.  Since $\mathcal R$ is strict, this means that every
infinite multipath has an infinitely-descending thread, hence SCT is satisfied.

Suppose to the contrary that there is a counter-example $M^R$.
In fact, let $M^R$ be a shortest counter-example. Then $M^R$ ends in
a graph $G^R_c$ and there is a thread $\tau$
leading from $\vec v_0$ at the front of $M^R$ up to a node
$\vec v$ on the source side of $G^R_c$
such that $\nxt(\vec v, G_c)$ is undefined.

Let us review the cases in which $\nxt(\vec v, G_c)$
is undefined:
\begin{enumerate}[(1)]
\item
$|\Im_1(\vec{v},G)| = 0$.
\item
The first descending position is $i$ and $|\Im_i(\vec{v},G)| = 0$ and
$v_{i-1}=0$.
\item
No position is descending and the vector ends with a 0.
\end{enumerate}

We next undertake to derive a contradiction in each of these cases.
\paragraph*{Case 1.}
This case implies that there is no complete thread in $M^G$;
indeed, when looking at the vectors of $\tau$, we see that initially all the variables are in 
the lowest position, and all threads must stay there
(review the definition of $|\Im_i(\vec{v},G)|$); by the definition of this case,
none of them survives to the end of the multipath.

Since $\mathcal G$ is strongly connected, it is possible to complete $M^G$ by
adding size-change graphs, if necessary, so that the multipath ends at $f_0$.
We thus obtain a cycle without any complete thread, which
contradicts the SCT property of $\mathcal G$ (repeat the cycle forever and
obtain an infinite multipath without infinite descent).

\paragraph*{Case 2.}
$|\Im_i(\vec{v},G_c)| = 0$, $v_{i-1}=0$ while
for all odd $j<i$, $|\Im_j(\vec{v},G_c)| = |v_j|$ and there are only non-strict
arcs between $v_j$ and $\Im_j(\vec{v},G_c)$.

Observe that $v_{i-1}$ can only reach 0 by being decreased repeatedly, from 
the last point where it had
a value of $B$ down to zero; this means that the
$i$-band of $\tau$ is stable.

Informally speaking, we can ignore the $i$-band and look at the vectors starting
at the $i$th position, and there we will find a situation similar to Case~1. 
For this reasoning to be correct, it is necessary
that the set of variables within the $i$-band be
the same at the beginning and the end of the ``bad'' multipath. We will use a 
pigeon-hole argument to show that such a situation must occur.

Consider the steps $\vec v_{t-1}\stackrel{G_t}{\to}\vec v_t$ 
of $\tau$ where position $i-1$ is decremented.
Label each such step with its target flow-point, say $f$, and 
the subset $S_1 = P_{i-1}(\vec v_t) \subseteq \parm(f)$.
The value $B=m\cdot 2^n$ guarantees that some pair $(f,S_1)$
must occur twice. Choose two positions $s<t$ that get the same label $(f,S_1)$. 
Then $\nxt(\vec v_{s-1}, G_s)$
as well as $\nxt(\vec v_{t-1}, G_t)$ are defined by (N2). Therefore,
$\vec v_{s}$ has
 $S_2 = \parm({f}) \setminus S_1$ in position $i$. In addition,
$|\Im_i(\vec{v}_{t-1},G_t)| = 0$, that is, the set in position $i$
has no outgoing arcs in $G_t$.

Now, consider the $\mathcal G$-multipath $C = G_{s+1}\dots G_t$,
 starting and ending at $f$ (Figure~\ref{fig-Case2}). The set
$S_1$ is carried by non-strict threads of $C$ unto itself, while
no complete thread starts and ends within $S_2$.
Conclusion: repeating this cycle infinitely many times, there will
be no infinite descent, contradicting the termination of $\mathcal G$.

\paragraph*{Case 3.}
No position in $\vec v$ is descending and the vector ends with a 0.

This case can be seen as a sub-case of Case 2, where $S_1$ includes
all variables.
\eprf

\bthm \label{thm-fof}
Suppose that ${\mathcal G}$ is a fan-out free, positive SCT instance. Let $B=m\cdot 2^n$.
There exists an indexed set $\{S_f\}$, where $S_f \subseteq V_f^B$
for every $f\in F$, such that the function
$\rho_f(\sigma) = \minval{\sigma}{S_f}$
is a ranking function for $\mathcal G$.
This ranking function can be effectively constructed given $\mathcal G$.
\ethm

\bprf
Since $\mathcal R$ is all-strict, fan-out free, and satisfies SCT, 
Theorem~\ref{thm-min} applies. By constructing $\mathcal R$ and computing
its maximal thread preserver, we find the
sets of vectors whose minima comprise the ranking function.
\eprf

Practically, $\mathcal R$ need not be constructed to its full doubly-exponential size;
see Section~\ref{sec:fof-example}.

\subsection{A useful observation} \label{SMOdag}
 We note a fact that will prove useful
in the sequel.  Suppose that we modify the SMO into a partial order 
SMO$^\dagger$ by
defining sets of different cardinalities to be incomparable. We can still
carry out our construction. The trick is simple: if we replace a set $S$
by the pair $\langle |S|, S\rangle$, the intended order is restored. An extra
numeric position before a set-valued position $v_i$ is not actually needed,
 because it can be merged into the numeric position $v_{i-1}$ 
(replacing $\langle v_{i-1}, |v_i|\rangle$ with $v_{i-1}n + |v_i|$;
the upper bound $B$ changes into $Bn$). The case of $v_1$ is different: here
the prefix $|v_1|$ is not necessary at all, because in an infinite thread,
the size $|v_1|$ must be eventually constant.

\subsection{The size of the ranking function}

If the size of the ranking function (more precisely, the expression for that
function) is of interest, 
the expression that results of the previous construction
should be optimized by eliminating redundancies, as
explained next.

\bdfn
Let $\vec v, \vec u \in V_f$. The relation $\vec v \prec \vec u$ holds
if there is an index $i$, such that for all $j<i$, $v_j=u_j$, while
$|v_i|<|u_i|$.
\edfn

\begin{obs}\label{obs:domination}
If $\vec v \prec \vec u$ then, regardless of the values of variables,
the value of $\vec v$ is lexicographically smaller than the value of
$\vec u$. We say therefore that $\vec v$ \emph{dominates} $\vec u$.
\end{obs}

Clearly, if there are dominated vectors in the ranking expression, they
can be dropped. A set without dominated vectors is said to be
\emph{in normal form}.

\begin{lem} \label{lem-n!}
Any set $R\subseteq V_f$ in normal form satisfies $|R|\le \atseret{n}$.
\end{lem}

\bprf
Observe that all first positions of vectors $\vec v\in R$ must
contain sets of the same size $k_1$.
We prove the lemma by induction on $n-k_1$.
If $k_1=n$ then there is just one set
and clearly $|R|=1$. If $k_1<n$ there are at most ${n\choose k_1}$ different sets
in the first position.
Given any choice $S_1$ for the first position, the number in the second position
is unique (by normality). If we choose from $R$ just the vectors beginning
with $S_1$ and drop the first two positions, we obtain a normal set over
$n-k_1$ variables. By the induction hypothesis, its size is bounded by
$(n-k_1)!$ and the bound on $|R|$ readily follows.
\eprf

\begin{cor} \label{cor-fof}
Every fan-out free, terminating SCT instance has a ranking function
of form $\rho_f(\sigma)=\minval{\sigma}{S_f}$
with $S_f\subseteq V_f$ in normal form. In particular, $|S_f|\le  \atseret{n}$.
\end{cor}

We can also simplify the form of the function---expressing it without the
use of multisets.
Let $\widetilde V_f^B$ be the set of $V_f^B$ vectors in which all odd positions 
are singletons. When working exclusively with such vectors, the multiset
concept is redundant. One can define the value of a position
holding $\{x\}$ as $\sigma(x)$ and consider
the vectors to have values in $(\val\times\nats)^n$.

For $\vec{v}\in V_f$, a {\em simplification} of $\vec{v}$ is obtained
as follows. 
For every odd position $i$, replace the set $v_i$ with 
a list of its elements (if $|v_i|>1$, there are different ways to do that,
so there are many simplifications).
Insert $0$ between any pair of consecutive non-numerical positions,
to obtain an element of $\widetilde V_f^B$.
For $R\subseteq V_f$, let $\widetilde R$ be the set of all simplifications of
$\vec{v}\in R$.
 
\begin{lem}
Suppose that $S_f\subset V_f^B$ and $S_g\subset V_g^B$ are normal.
If $\minval{\sigma}{S_f} > \minval{\sigma'}{S_g}$, then also
$\minval{\sigma}{\widetilde S_f} > \minval{\sigma'}{\widetilde S_g}$.
\end{lem}

\bprf 
Let $\widetilde{\vec v}\in \widetilde S_f$ be the vector of minimal value
(under $\sigma$). Consider $\vec v$; the assumption 
 $\minval{\sigma}{S_f} > \minval{\sigma'}{S_g}$ implies that there is
 a vector $\vec u\in S_g$ such that $\sigma'(\vec u) < \sigma(\vec v)$.
Note that this is a lexicographic comparison. Thus there is an index $i$
such that for all $j<i$, $\sigma'(\vec u)_j = \sigma(\vec v)_j$,
while $\sigma'(\vec u)_i < \sigma(\vec v)_i$.

The equalities at positions $j<i$
mean that for all the sets in such positions, the elements of $u_j$
can be arranged to match elements of $v_j$ of equal value as they appear
in $\widetilde{\vec v}$. Thus we build a simplification $\widetilde{\vec u}$
whose value is identical to that of
$\widetilde{\vec v}$ up to position $i$. If this
position is numeric, then we can clearly complete the simplification and
obtain $\sigma(\widetilde{\vec v}) > \sigma'(\widetilde{\vec u})$.
If position $i$ is a set, the multiset
inequality $\sigma'(\vec u)_i < \sigma(\vec
v)_i$ means that, if the elements of both vectors are arranged in
ascending order of value, the vector obtained from $u_i$ will be
lexicographically smaller. Observe that among all listings of the elements
of a multiset, the ascending list \emph{is} lexicographically smallest.
Therefore, the elements of $v_i$ must appear in $\widetilde{\vec v}$
in exactly that order (as it is minimum in $\widetilde S_f$). Now,
arranging the elements of $u_i$ in corresponding order, we obtain a
simplification $\widetilde u$ that is smaller than $\widetilde v$.

Thus some element of $\widetilde S_g$ has smaller value than $\widetilde v$,
which shows that $\minval{\sigma'}{\widetilde S_g}$ is definitely smaller than
$\widetilde v$.
\eprf

\begin{thm} \label{thm-fofsimplified}
Every fan-out free, positive SCT instance $\mathcal G$ has a ranking function
of form $\rho_f(\sigma) = \minval{\sigma}{S_f}$ with
$S_f\subseteq \widetilde V_f^B$ and $|S_f|\le  \atseret{n}$.
\end{thm}

\bprf 
We start with a ranking function as claimed in Theorem~\ref{thm-fof}
and replace every set $S_f$ by its simplification; then we remove dominated vectors
to obtain a normal set.
\eprf

\subsection{When the CFG is not strongly connected} 

For completeness, let us explain how ranking functions of the kind
constructed in this paper should be adjusted when the control-flow graph
is not strongly connected.

Suppose that the CFG of ${\mathcal G}$ consists of several strongly connected
components (SCCs). Let $C_1,\dots,C_k$ be a reverse topological ordering
of the components.
If the size-change graphs in every component are fan-out (or fan-in) free,
our constructions yields a function $\rho_f(\sigma)$ 
that decreases in every transition within a component (note that
we construct the function separately for each component,
but it can be considered as one function; no conflicts arise because
every flow-point $f$ belongs to a unique component).

Next, define $\rho'(s)$ for any state $s = (f,\sigma)$, where $f\in C_i$,
by prepending $i$ to $\rho_f(\sigma)$ (this gives a vector with an extra
numeric position at its beginning).
 It is easy to see that $\rho'$ is a ranking function for ${\mathcal G}$.

\subsection{An example, and a simplification of the algorithm} \label{sec:fof-example}

\newcommand{\xx}{{\tt x}}
\newcommand{\yy}{{\tt y}}
\newcommand{\zz}{{\tt z}}

Next, we carry out the construction for a very small example. Small, but hopefully illustrative.
In fact, the example will illustrate how the construction can be made
simpler and much more efficient than a literal implementation of the proof of 
Theorem~\ref{thm-fof}. After explaining the example, we will also formulate the simplification in
 general terms.

Our example is an instance with $F=\{f\}$, $\parm(f) = \{\xx,\yy\}$
 and ${\mathcal G}=\{G_1, G_2\}$, where
$$
G_1 = \{ \xx \stackrel{\downarrow}{\to} \yy ,\; \yy \stackrel{\downarrow}{\to} \yy \} ,\;
G_2 = \{ \xx \stackrel{\downarrow}{\to} \xx ,\; \yy \stackrel{\deqsm}{\to} \xx \} \,.
$$
We have $B= m\cdot 2^n = 4$.
We will not construct $\mathcal R$ in full, which would have meant (if we followed
Definition~\ref{def:R} literally) creating an instance with $|V_f| = 36$ variables.
Instead, we note that according to the proof of Theorem~\ref{thm:R}, it suffices
to find the vectors reachable from ${\vec v}_0 = \langle \parm({f_0}), B\rangle$
 by applications of $\nxt$; this will
be a smaller sub-instance of $\mathcal R$ which satisfies the theorem.

So, the actual procedure is as follows: we initialize a collection $S$ of vectors to
$\{ {\vec v}_0 \}$. We repeatedly compute, for each vector $\vec v  \in S$,
and $G_i\in {\mathcal G}$, the result of $\nxt(\vec v, G_i)$, adding it to $S$, until $S$
does not grow any further. We describe the result of this process as a graph,
where the arcs out of $\vec v$ describe
the applications of $\nxt(\vec v, G_i)$ in the construction:

%in Figure~\ref{fig:fof-ex-graph}.

%fig0 causes "TeX capacity exceeded" message!!
%\begin{fig0}{Closure of $\{ {\vec v}_0 \}$ under $\nxt(G_i,\vec v)$.}{t}{fig:fof-ex-graph}
$$
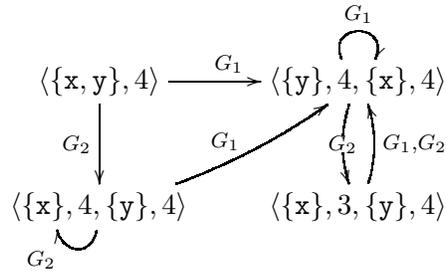
\xymatrix@R=30pt{
    {\langle \{\xx,\yy\} , 4\rangle}  \ar[d]_{G_2}  \ar[r]^{G_1} &
     {\langle \{\yy\} , 4, \{\xx\} , 4\rangle}  \ar@(ul,ur)[]^{G_1} \ar@/_/[d]|{G_2} \\
     {\langle \{\xx\} , 4, \{\yy\} , 4\rangle}  \ar@(d,l)[]^{G_2}  \ar@/_/[ru]^{G_1} &
     {\langle \{\xx\} , 3, \{\yy\} , 4\rangle}  \ar@/_/[u]_{G_1,G_2} \\
}$$
%\end{fig0}

The set of all four vectors is a thread preserver,
as each of them has an outgoing arc both under $G_1$ and under $G_2$. 
 We conclude that
$$\rho(\xx,\yy) = \min(
{\langle \{\xx,\yy\} , 4\rangle} , \;
     {\langle \{\yy\} , 4, \{\xx\} , 4\rangle}  , \;
     {\langle \{\xx\} , 4, \{\yy\} , 4\rangle}  , \;
     {\langle \{\xx\} , 3, \{\yy\} , 4\rangle} 
  )$$
is a ranking function. Removing dominated vectors, we reduce the expression to
$$\rho(\xx,\yy) = \min(
     {\langle \{\yy\} , 4, \{\xx\} , 4\rangle}  , \;
     {\langle \{\xx\} , 3, \{\yy\} , 4\rangle} 
  ).$$
  
The fact that the set of vectors constructed, $S$, constituted a thread preserver,
is not an incident. We next demonstrate this in general. This means that the way of
computing $S$ not only economizes
on the size of $\mathcal R$, it also obviates the need for an MTP computation. 

\bthm
Let $S\subseteq V_f$ be the closure of the set $\{\vec v_0\}$, where $\vec v_0 = \langle \parm({f_0}), B\rangle$,
under the operators $\nxt (\vec  v, G)$ for all $G\in \mathcal G$. This set is a thread preserver in
$\mathcal R$.
\ethm

\bprf
By the definition of $S$, if $\vec v \in V_f$ is in $S$, this is because
 there exists a sequence
$\vec v_0, \vec v_1, \dots, \vec v_k = \vec v$ such that for all $i$,
$\vec v_i = \nxt(\vec v_{i-1}, G_i)$ for an appropriate $G_i$. Thus, in the
$\mathcal R$-multipath $M = G_1 G_2\dots G_k$, vector $\vec v$ lies on a thread $\tau$ emanating
from $\vec v_0$. The crucial observation now is that, \emph{since $\mathcal R$ is fan-out free,
there is just one such thread}. This fact was not used in the proof of Theorem~\ref{thm:R},
but it means that we can deduce from the proof that this particular thread $\tau$ can be
continued in any $\mathcal R$-multipath extending $M$, say $MG_{k+1}$. Thus, for any
applicable $G_{k+1}$, $\nxt(\vec v, G_{k+1})$ is defined---and has to be in $S$.

We conclude that $S$ has the property of a thread preserver.
\eprf

This technique can be pushed a little further, as shows the next theorem.
 
\bthm
Considering $S$ as the node set of a directed graph $(S,A)$ 
(with arcs $\vec v \to \nxt(\vec v, G)$),
let $C$ be a sink SCC of this graph. Then $C$ is a thread preserver.
\ethm

\bprf
The thread-preservation follows directly from the definition of $S$ plus
the fact that no arc leaves a sink SCC.
\eprf

Thus, a standard SCC algorithm (which runs in linear time~\cite{CLRS}) suffices for
obtaining a thread preserver, which can be smaller than the MTP, since it is actually a
\emph{minimal} thread preserver (as the interested reader may verify). In the last example,
this TP consists of the two vectors that wound up in the final ranking function. 
It is not true in general, however, that the SCC will be a normal set (free of dominated vectors),
as was the case with this example. A counter-example is given by ${\mathcal G}=\{G_1, G_2\}$, where
$$
G_1 = \{ \xx \stackrel{\downarrow}{\to} \xx ,\; \yy \stackrel{\deqsm}{\to} \yy ,\;
 \zz \stackrel{\downarrow}{\to} \yy\} ,\;
G_2 = \{ \yy \stackrel{\downarrow}{\to} \zz ,\; \zz \stackrel{\downarrow}{\to} \zz \} \,.
$$

\paragraph{An additional implementation tip}
The value of $B$ used in Theorem~\ref{thm-fof}, $m\cdot 2^n$, becomes unwieldy if $n$ is large.
In fact, it is an overestimate. Even theoretically, the range 0 to $B$ will never be fully
used (the interested reader is invited to prove it), and in most cases a much smaller
range will be needed. The right way to implement the numeric positions is by 
inverting the interval $[0,B]$, so that they are initially 0 and increasing, instead of
starting at $B$ and decreasing.
For preserving the natural order on numbers, invert again once the true
range necessary for an instance has thus been discovered.

\paragraph{Other optimizations}
The reader may have noticed that the function:
$$\rho(\xx,\yy) = \min(
     {\langle \{\yy\} , 4, \{\xx\} , 4\rangle}  , \;
     {\langle \{\xx\} , 3, \{\yy\} , 4\rangle} 
  )$$
can be simplified to
$$\rho(\xx,\yy) = \min(
     {\langle \{\yy\} , 4\rangle}  , \;
     {\langle \{\xx\} , 3\rangle} 
  )$$
while remaining a ranking function. We have not generalized this observation or formulated
a procedure to find such savings.

\section{Fan-in Free SCT} \label{sec-fif}

The fan-in free class of SCT instances is symmetric in nature to the
fan-out free class. This symmetry has been used in~\cite{BL:2006} where
it was also observed that transposing graphs, which clearly makes fan-in
into fan-out (and vice versa), also has to do with exchanging {\bf
min}-descent with {\bf max}-descent; this is illustrated in the simple results
cited in Section~\ref{sec-strict}.

In this section, we show how to use transposition for applying
the construction and results of Section~\ref{sec-fof} to fan-in free graphs.
The arguments here are somewhat subtler than those of Section~\ref{sec-strict},
since they involve the semantic connection between $\mathcal G$ and $\mathcal R$.
To do this precisely, we have to review some of our concepts and results
with an eye to more generality. 

We begin by noting that the SCT condition (Section~\ref{sec:SCT})
is completely independent
of the semantic interpretation of size-change graphs, involving the order
relation on $\val$, to which we now assign the notation $\le_{\val}$.
This order is relevant however for defining the transition system 
$T_{\mathcal G}$ (Definition~\ref{def-transitionsystem}), so for clarity we may
notate it as $T_{\mathcal G}[\le_{\val}]$. Note that the assumption that the order
is well-founded is only necessary for justifying the conclusion regarding
termination of $T_{\mathcal G}$.
In particular, the construction of $\mathcal R$ is, obviously, completely syntactic,
while the justification of its size-change arcs can be put in more general terms
as follows.

\newcommand{\vectorset}[2]{\mathrel{(#1  \otimes^n #2)}}
\newcommand{\vecorder}[2]{\mathrel{({#1}  \otimes^n {#2})}}
\newcommand{\vecordernp}[2]{{#1 \; \otimes^n #2}} %no paren's

For a given order $\le_{\val}$, let $\mbox{SMO}^\dagger[\le_{\val}]$ denote
the simple multiset partial order (as described in Section~\ref{SMOdag})
parameterized by the order $\le_{\val}$ for set elements.

\bdfn
For given orders $\le_{\val}$ and $\le_B$ (with carrier sets $\val$ and
$[0,B]$, respectively), let $\vectorset{\val}{B}$ denote the set of
tuples $\langle v_1,v_2,\dots\rangle$ of even length,
where every odd position is a non-empty multiset over $\val$, 
together containing $n$ elements;
and every even position is an integer in $[0,B]$. The lexicographic
partial order on $\vectorset{\val}{B}$, obtained by ordering odd positions
with $\mbox{SMO}^\dagger[\le_{\val}]$
and even positions with $\le_B$, is denoted by
$\,\vecordernp{\le_{\val}}{\le_B}$.
\edfn

The following is a parametrized rereading of Observation~\ref{obs-next}:

\begin{claim} \label{clm-R}
For a size-change graph $G$ and a vector $\vec v$,
$$((f,\sigma),(g,\sigma'))\in T_{\mathcal G}[\le_{\val}]
\;\Longrightarrow\;
\sigma'(\nxt(\vec v, G)) \;\vecordernp{\le_{\val}}{\le_B}\; \sigma(\vec v)
\,.$$
\end{claim}

This shows that the interpretation of $\mathcal R$ as size-change graphs works
in the general setting.

We now move to transposition. 
In Section~\ref{sec-strict}, we observed that the SCT property of $\mathcal G$
is preserved under transposition. But what do the transposed graphs describe?
The natural answer is given by the equation
$$T_{{\mathcal G}^t}[({\le_{\val}})^{t}] = (T_{\mathcal G}[\le_{\val}])^{t}$$
(a transposed relation is defined in the natural way; $\le^{t}$ is
the reverse order relation, $\ge$).

Here is a useful lemma concerning transposition.
\blem\label{lem:translex}
$(\vecordernp{(\le_{\val})^{t}}{\le_B})^{t} = \vecordernp{\le_{\val}}{\,(\le_B)^{t}}$.
\elem

We leave its proof to the reader;  note that using the
partial version SMO$^\dagger$ helps to avoid the asymmetric definition of SMO
 for sets
of different size.

\bthm \label{thm-fif}
Let ${\mathcal G}$ be fan-in free, positive SCT instance, and let $B=m\cdot 2^n\cdot n$.
There exists an indexed set $\{S_f\}$, where $S_f\subseteq V_f^B$
for every $f\in F$, such that the function
$\rho_f(\sigma) = \maxval{\sigma}{S_f}$
is a ranking function for $\mathcal G$.

This ranking function can be effectively constructed given $\mathcal G$.
\ethm

\bprf
Observe that this is a version of Theorem~\ref{thm-fof}, with $\minval{}{}$ changed
into $\maxval{}{}$ (this will be justified shortly) and $B$ multiplied by $n$
to compensate for the use of SMO$^\dagger$ instead of SMO (see 
Section~\ref{SMOdag}).
Now, we describe the construction of $\rho$.

Given $\mathcal G$, construct ${\mathcal G}^t$; note that will be fan-out free.
Thus $\mathcal R$ can be constructed from it as in the last section, and it is
fan-out free and strict. 
Since the interpretation of ${\mathcal G}^t$ uses the ordering ${\le_{\val}}^{t}$,
the semantics of $\mathcal R$ is given by a transition system
$$T_{\mathcal R}[\vecordernp{(\le_{\val})^{t}}{\le_B}].$$
Now, ${\mathcal R}^t$, is
strict, fan-in free, and interpreted under the order
$(\vecordernp{(\le_{\val})^{t}}{\le_B})^{t}$.

Since $\mathcal R$ satisfies SCT (as proved in the last section), so does
${\mathcal R}^t$, and by Theorem~\ref{thm-max} it has a ranking function $\rho$
of the form stated in the current theorem. This function decreases under
the ordering $(\vecordernp{{(\le_{\val})}^{t}}{\le_B})^{t}$. 

Since by Lemma~\ref{lem:translex},
$(\vecordernp{{(\le_{\val})}^{t}}{\le_B})^{t} = \vecordernp{\le_{\val}}{\,(\le_B)^{t}}$,
we find that $\rho$ descends under the usual ordering of $V_f^B$, except that
the numeric positions are ordered in reverse (${\le_B}^{t}$ instead of $\le_B$).
But this can be easily fixed by exchanging every value $v_{2i}$ by $B-v_{2i}$,
so that descent in $\vecorder{\le_{\val}}{\le_B}$ is obtained.
\eprf

The simplifications considered in the previous section also apply here.

\begin{cor} \label{thm-fifsimplified}
Every fan-in free, positive SCT instance has a ranking function
of form $\rho_f(\sigma) = \maxval{\sigma}{S_f}$ with
$S_f\subseteq \widetilde V_f^B$ and $|S_f|\le  \atseret{n}$.
\end{cor}

\section{Lower Bounds}

Our upper bound on the size of the ranking functions is exponential; more precisely,
up to $n!$ vectors under the \textbf{min} or \textbf{max} sign, for every flow-point.
What is the true complexity? 
In this section we provide (mostly) explicit lower bounds, that is, lower bounds on
the ``size" of \emph{any} ranking function for a specific family of
SCT instances. In order to prove such a result,
it is necessary to make
assumptions on the form of the ranking function.  We progress through three types
of ranking functions that generate vectors of variables and constants, as our
constructions do. The first type precludes the use of multisets, the second allows
them with a restriction, and the third is the most general.

All our examples are for a CFG with a single node $f$.
We consider a class of ranking functions that can be described by assigning an
element $\vec v \in V_f^B$ to any
total order ${\tau}$ on $\parm(f)$, 
such that if the values of variables in state $s$ satisfy $\tau$, then
$\rho_f(s)$ is given by $\vec v$ (we can ignore the value of the
function in the case that some variables tie).
The value of $B$ is left unspecified.
Note that
the constructions in this paper are of this kind, since if the order of variable values is known,
the lexicographic or multiset minima and maxima can be deduced.
We call such ranking functions VSO (for Vectors Selected by Order).

\bdfn 
Let $s=(f,\sigma)$ be a state, and $\pi\in S_n$
an $n$-element permutation.
 We say that a state $s$ has order $\pi$ whenever
$\pi i < \pi j \iff s(x_i) < s(x_j)$ for all $i,j$. If the values of variables in $s$ are distinct,
$\pi$ is unique and we denote it by $\mbox{\it Order}(s)$.
\edfn

\bdfn Let $\rho(f,\sigma)$ (or $\rho_f(\sigma)$) be a function over
states with co-domain $V_f^B$. Such a function is called a \emph{VSO
function} if there is, for each $f$, a function $\rho_f^* : S_n \to
V_f^B$ such that, for all $s=(f,\sigma)$ where the values of all
variables are distinct, we have $\rho_f(\sigma) = s(\rho_f^*(\mbox{\it
Order}(s)))$ (recall that $s(\vec v)$ is the value of vector $\vec v$
in state $s$).  \edfn

We knowingly disregard the states where variable values are not distinct.

The lower bounds in this section apply to the number of distinct vectors in the image set
of $\rho_f^*$, i.e. to the size of the set $\rho_f^*(S_n)$.
Thus they apply to the number
of vectors under the $\min$ or $\max$ operators in functions expressed as in
Theorems~\ref{thm-fof} and~\ref{thm-fif}.
All our constructions are fan-out free instances, and the lower bounds are very close to
the upper bounds we had for this class.

We believe that in a certain sense, a ranking function generated by 
a general SCT-based construction (i.e., not using any other information, say about $\val$)
\emph{has} to be a VSO function (or representable as one)
since all that
is assumed of the data $\val$ is that they are ordered, so decisions can only be based
on order; and variable values can be put into tuples or multisets, but not otherwise
used in expressions. We have not formalized this intuition, however.

\subsection{Preliminaries}

The set $\parm(f)$ is written in the following examples as $\{x_0,x_1,\dots,x_n\}$ (so there are
actually $n+1$ variables).   States may be written in the form $[x_0\mapsto v_0,\dots,
x_n\mapsto v_n]$.
In our examples, values are non-negative integers, and
$x_0$ is always largest. 
We also assume that all variable values are different. 
Therefore, the ordering of variable values in a state will be described by a permutation on 
$[1,\dots,n]$, reflecting the ordering among $x_1,\dots,x_n$.

We denote by $I$ the identity on $[1,\dots,n]$
and by $\xi_{ij}$ the permutation exchanging $i$ and $j$. 
 Composition of permutations is defined by the rule
$(\pi_1 \pi_2) i = \pi_1 (\pi_2 i)$.  For a state $s$, $s\pi$ denote the result of the action
of $\pi$ on the state $s$, defined by $(s\pi)(x_i) = s(x_{\pi i})$.

\subsection{Simple ranking functions}

We observed (in Section~\ref{sec-fof})
 that it is possible to restrict all our constructions to vectors which
do not employ multisets (technically, all odd positions are singletons).
A ranking function of this type is \emph{simple}.
We begin by proving a lower bound for simple ranking functions.

\begin{thm} \label{thm-lb1}
Let $n>0$.
There is a fan-out free positive SCT instance $\mathcal G$ with a single flow-point,
$n+1$ variables and $n$ size-change graphs 
such that any simple VSO ranking function $\rho_f$ for $\mathcal G$ satisfies
 $|\rho_f^*(S_n)| \ge n!$.
\end{thm}

Observe that this lower bound matches almost exactly the upper bound of 
Theorem~\ref{thm-fofsimplified}.

The proof of the theorem breaks into the following parts: (1) construction of the SCT
instance; (2) proof that SCT is satisfied; (3) proof of the lower bound.

\subsubsection*{Construction {\thesection}.1: SCT instance $\mathcal G$}
{\abovedisplayskip=\abovedisplayshortskip
Let ${\mathcal G} = \{G_1,\dots,G_{n}\}$, where for $k=1,\dots,n-1$,
$$\begin{array}{rcl}
G_k &=& \{ x_k \to x_{k+1}, x_{k+1}\to x_k \} \cup \{ x_i\to x_i \mid i\neq k,k+1\} 
         \cup \{ x_0 \stackrel{\downarrow}{\to} x_0 \}
\quad\mbox{and\mathstrut}\\
G_n &=& \{ x_i \to x_i \mid 1\le i < n \}
         \cup \{ x_n \stackrel{\downarrow}\to x_n \}.
\end{array}$$
(See Figure~\ref{fig-calG} for an illustration.)}

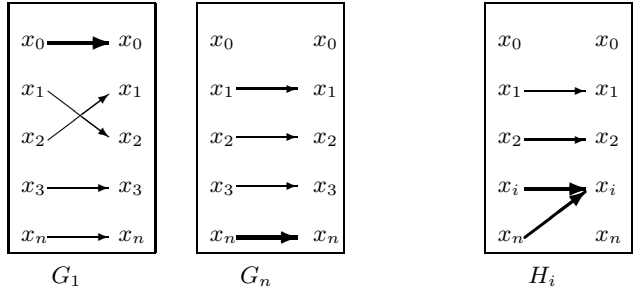
\begin{fig1}{A partial view of two graphs from Construction {\thesection}.1
and one from Const.~{\thesection}.2.}{t}{fig-calG}
{\setlength{\unitlength}{0.46pt}\footnotesize
\begin{center}
\begin{picture}(300,240)(100,-55)
	      \put( 85,-20){\framebox(120,205){}}
             \put( 95,150){${x}_0$}
			 \put( 95,110){${x}_1$}
              \put( 95, 70){${x}_2$}
              \put( 95, 30){${x}_3$}
              \put( 95, -10){${x}_n$}
              
	      \thinlines
	      \put(117,33){\vector(1,0){50}}
	      \put(117, -7){\vector(1,0){50}}
	      
	      \thicklines
	      \put(117,153){\vector(1,0){50}}
	      \put(117,152){\vector(1,0){50}}
	      \thinlines
	      
	      \put(117,73){\vector(4,3){50}}
	      
	      \put(117,113){\vector(4,-3){50}}
	      
\thinlines
             \put(175,150){${x}_0$}
              \put(175,110){${x}_1$}
              \put(175, 70){${x}_2$}
              \put(175, 30){${x}_3$}
              \put(175, -10){${x}_n$}

	      \put(240,-20){\framebox(120,205){}}
              \put(250,150){${x}_0$}
               \put(250,110){${x}_1$}
               \put(250, 70){${x}_2$}
               \put(250, 30){${x}_3$}
              \put(250, -10){${x}_n$}

	      \put(272,115){\vector(1,0){50}}
          \put(272,75){\vector(1,0){50}}
	      \put(272,35){\vector(1,0){50}}
	      
	      \thicklines
	      \put(272,-7){\vector(1,0){50}}
	      \put(272,-8){\vector(1,0){50}}
	      \thinlines
             \put(335,150){${x}_0$}
             \put(335,110){${x}_1$}
             \put(335, 70){${x}_2$}
             \put(335, 30){${x}_3$}
             \put(335, -10){${x}_n$}

	      \put(120,-45){$G_1$}
	      \put(275,-45){$G_n$}
\end{picture}
\hspace{0.5in}
\begin{picture}(200,240)(100,-55)
	      \put( 85,-20){\framebox(120,205){}}
             \put( 95,150){${x}_0$}
			 \put( 95,110){${x}_1$}
              \put( 95, 70){${x}_{2}$}
              \put( 95, 30){${x}_i$}
              \put( 95, -10){${x}_n$}

	      \thicklines
	      \put(117,33){\vector(1,0){50}}
	      \put(117,32){\vector(1,0){50}}
	      \put(117,-7){\vector(4,3){50}}
	      \put(117,-8){\vector(4,3){50}}
	      \thinlines
	      
	      \put(117,73){\vector(1,0){50}}
	      
	      \put(117,113){\vector(1,0){50}}
	      
\thinlines
             \put(175,150){${x}_0$}
              \put(175,110){${x}_1$}
              \put(175, 70){${x}_{2}$}
              \put(175, 30){${x}_i$}
              \put(175, -10){${x}_n$}
	      \put(120,-45){$H_i$}
\end{picture}
\end{center}
}
%\vair
\end{fig1}

%%%%%%%%%%%%%%%%%%%%%
\bigskip

Observe that any contiguous sequence of the graphs $G_1,\dots,G_{n-1}$
represents (or ``effects'') some permutation $\pi$ on $\{x_1,\dots,x_{n}\}$, in the sense that
every $x_i$ is carried by a (non-descending) thread to $x_{\pi i}$.
Henceforth, we call this ``a permutation multipath'' $M_\pi$.
Note that for every permutation $\pi\in S_n$,
there is a multipath effecting it, less than $n^2$ long, according
to well-known ways of composing any permutation of exchanges.

\blem
$\mathcal G$ satisfies SCT.
\elem

\bprf
In any infinite $\mathcal G$-multipath that does not contain $G_n$ (or where $G_n$ occurs
a finite number of times), there is infinite descent at $x_0$.
In a multipath that includes an infinity of $G_n$'s, the thread at 0 is lost
but each of $x_1,\dots,x_n$ begins a separate infinite thread.
At each occurrence of $G_n$, one of these threads descends,
so there is at least one thread of infinite descent.
\eprf

\blem \label{lem-0last}
Let $\rho_f$ be a simple ranking function for $\mathcal G$ and $\vec v = \rho_f^*(\pi)$
for some permutation $\pi$.
Then $x_0$ is the last variable that appears in $\vec v$.
\elem

\bprf
Consider a state $s$ with $\mbox{\it Order}(s) = \pi$,
so $\rho_f(s) = s(\vec v)$. Let $i>0$. 
Let $s'$ be identical to $s$ except that $x_i$ decreased and $x_0$
increased, while the relative order of variables remains $\pi$, so $\rho_f(s')$ is also
given by $\vec v$.
Note that $M_{\xi_{in}} G_n M_{\xi_{ni}} \models s\mapsto s'$. 
Thus, we must have $\rho_f(s)>\rho_f(s')$, so $x_i$ must appear in $\vec v$ before $x_0$.
Since this conclusion holds for every $i>0$, it follows that $x_0$ must be last.
\eprf

\newcommand{\plist}[1]{{\mbox{\sc LIST}(#1)}}
\newcommand{\plistvec}[1]{{L_{#1}}}

For any $\vec v$, let $\plist{\vec v}$ denote the permutation that describes the placement
of variables in $\vec v$ (excluding $x_0$ which we know to be last). That is,
$\plist{\vec v}i = j$ indicates that the $i$th variable occuring in $\vec v$ is $x_j$.
For a permutation $\pi$, we use the following abbreviations:
$\vec v_\pi$ is $\rho_f^*(\pi)$; $\plistvec{\pi}$ is $\plist{\vec v_\pi}$.

\blem \label{lem-lb1}
Let $\rho_f$ be a simple ranking function for $\mathcal G$ and $\pi,\tau\in S_n$.
Then $\tau\plistvec{\tau} = \pi\plistvec{\pi}$.
Moreover, vectors $\vec v_\tau$ and $\vec v_\pi$ have the same constants
in the even positions below position $2n$.
\elem

\bprf
Assume the contrary. Consider the first position that violates the lemma.
Suppose first that it is an odd position $2i-1$, containing a variable.
Thus, letting $k = \plistvec{\tau}i$ and $j = \plistvec{\pi}i$,
we have $\tau k\ne \pi j$. Assume for the rest of the proof
that $\tau k > \pi j$ (otherwise, exchange $\tau$ and $\pi$).

Let $s=[x_j\mapsto \pi j \mbox{ for }1\le j \le n, x_0\mapsto n+n^2]$. Note that 
$\mbox{\it Order}(s) = \pi$. Thus, $\rho_f(s) = s(\vec v_\pi)$.
Let $s'=[x_k\mapsto \tau k \mbox{ for }1\le k \le n, x_0\mapsto n+1]$. Note that 
$\mbox{\it Order}(s') = \tau$. Thys, $\rho_f(s') = s'(\vec v_\tau)$.
It should be easy to see that $M_{\tau^{-1}\pi} \models s\mapsto s'$.

For all positions up to $2i-1$, the contents of $\rho_f(s)$ and $\rho_f(s')$ are
the same. In position $2i-1$, $\rho_f(s')$ has $s'(x_k) = \tau k$, while $\rho_f(s)$
has $s(x_j) = \pi j$. Now, 
By our assumption,
$\tau k > \pi j$, so $s'(x_k) > s(x_j)$.
We conclude that lexicographic descent fails, which contradicts  $\rho_f$ being
a ranking function.

Next, assume that the first position that violates the lemma
is an even position $2i < 2n$, containing a constant. The refutation is very similar.
\eprf

The proof of Theorem~\ref{thm-lb1} is now concluded since the last lemma implies
that a distinct vector corresponds to every permutation.

\subsection{Simply-ordered multisets}

Now we move to ranking functions with multisets of any size. 
In Section~\ref{sec-ms} we described forms of multiset ordering; 
our ranking functions are constructed so that the change in every multiset value
across a transition alway agrees with what we called \emph{the simple multiset
order} or SMO. This means that the relation between multisets of the same size
must be expressed by a 1--1 correspondence of the elements.
This subsection gives a lower bound under this restriction.

\begin{thm} \label{thm-lb2}
Let $n>0$.
There is a fan-out free positive SCT instance $\mathcal H$ with a single flow-point,
$n+1$ variables and $2n-1$ size-change graphs,
such that any SMO-descending VSO ranking function $\rho_f$ for $\mathcal H$ satisfies
$|\rho_f^*(S_n)| \ge n!$.
\end{thm}

Observe that this lower bound, too, closely matches the upper bound of Theorem~\ref{thm-fof},
though a larger class of functions is considered.
The proof proceeds through the same stages as the last one.

\subsubsection*{Construction {\thesection}.2: SCT instance $\mathcal H$}
{\abovedisplayskip=\abovedisplayshortskip
Let ${\mathcal H} = \{G_1,\dots,G_{n-1},H_1,\dots,H_n\}$, where
the $G_k$ graphs are identical to those of Construction {\thesection}.1, namely
$$G_k = \{ x_k \to x_{k+1}, x_{k+1}\to x_k \} \cup \{ x_i\to x_i \mid i\neq k,k+1\} 
         \cup \{ x_0 \stackrel{\downarrow}{\to} x_0 \}
$$         
while for $i=1,\dots,n$,
$$H_i = \{ x_j \to x_j \mid 0<j < i \}
         \cup \{ x_j \stackrel{\downarrow}\to x_i \mid j\ge i \}.
$$         
\par}

Thus, $\mathcal H$ contains $\mathcal G$ (note that $H_n$ is the same as $G_n$ of $\mathcal G$).

\blem
$\mathcal H$ satisfies SCT.
\elem

\bprf
In any infinite $\mathcal H$-multipath that does not contain $H$-graphs (or contains a finite
number of them), there is infinite descent at $x_0$.
In a multipath that includes an infinity of $H$'s, the set 
$\{x_1,\dots,x_n\}$ descends infinitely in dual multiset order,
%(Def.~\ref{def-ms}),
ensuring termination
(the restriction to SMO in Theorem~\ref{thm-lb2} only applies to our ranking function,
not to this proof!).
\eprf

\blem
Let $\rho_f$ be a SMO-descending VSO ranking function for $\mathcal H$ and
let $\vec v_\pi = \rho_f^*(\pi)$
for some permutation $\pi$.
Then $x_0$ only appears in $\vec v_\pi$'s last position.
\elem

\bprf
Essentially the same as for Lemma~\ref{lem-0last}.
Note that since $x_0$ increased from $s$ to $s'$,
we cannot obtain simple multiset descent in any set of variables that
includes $x_0$. Therefore, $x_i$ must appear strictly before $x_0$, not even in the
same set.
\eprf

\blem
Let $\rho_f$, $\pi$ and $\vec v_\pi$ as above.
Then every odd position of $\vec v_\pi$ is a singleton.
\elem

\bprf
The proof is by contradiction, as in Lemma~\ref{lem-lb1}.
Consider the first position that violates the lemma for some  $\pi \in S_n$.
Suppose that it is position $2i-1$.
Let $A$ be this element of $\vec v_\pi$, that is, the $i$th set-valued element.
Let $x_a$ be the variable in $A$ such that $b = \pi a$ is smallest.

Let $s=[x_j\mapsto 2\pi j \mbox{ for }1\le j\le n, x_0\mapsto n+n^2]$. Note that 
$\mbox{\it Order}(s) = \pi$. Thus, $\rho_f(s) = s(\vec v_\pi)$.
Let $s_1=[x_j\mapsto 2j \mbox{ for }1\le j \le n, x_0\mapsto n+1]$. Note that 
$\mbox{\it Order}(s) = I$, and that $M_{\pi}\models s\mapsto s_1$.
Let $$s_2=[x_j\mapsto 2j \mbox{ for }j<b,\, x_b\mapsto 2b-1,\,  x_j\mapsto 2n+j \mbox{ for }j>b, x_0\mapsto n+n^2].$$
Note that $H_b\models s_1\mapsto s_2$, and $\mbox{\it Order}(s_2)$ is  $I$.
Let $s_3 = (s_2\pi)[x_0 \gets n^2+1]$, so that $M_{\pi^{-1}}\models s_2\mapsto s_3$. 
We have
$\mbox{\it Order}(s_3) = \pi$ again. 

For all positions up to $2i-1$, the variables and constants in $\rho_f(s_3)$ have
at least their value as in $\rho_f(s)$.
In position $2i-1$, we have a set that includes the variable $x_a$,
and, if $|A|>1$, also other variables of greater indices. In $s_3$, those other variables
have values larger than in $s$. Thus, if lexicographic descent (under SMO)
is to be maintained, we must conclude that $|A|=1$.
\eprf

\blem
Let $\rho_f$ be a restricted ranking function for $\mathcal H$ and let $\pi, \tau\in S_n$.
Let $\vec v_\pi = \rho_f^*(\pi)$, and likewise $\vec v_\tau = \rho_f^*(\tau)$.
Then  $\tau\plistvec{\tau} = \pi\plistvec{\pi}$.
In addition, vectors $v_\tau$ and $v_\pi$ have the same constants
in the even positions before position $2n$.
\elem

\bprf
Given the last lemma,
the situation is identical to that in Lemma~\ref{lem-lb1} and the same proof holds.
\eprf

The proof of Theorem~\ref{thm-lb2} is now concluded since the lemma implies
that a distinct vector corresponds to every permutation.

\subsection{Dual-ordered multisets}

We now extend the allowable range of ranking functions further by allowing 
a stronger type of multiset ordering to be used. Since our example consists of
fan-out free graphs, it is easy to conclude that among the two orders described
in Section~\ref{sec-ms}, it is \emph{dual multiset order} (DMO) which is promising to be
useful. So, we consider VSO ranking functions in which multisets are compared
by DMO.

\begin{thm} \label{thm-lb3}
There is a fan-out free positive SCT instance $\mathcal K$ with a single flow-point,
$2n+1$ variables and $n+1$ size-change graphs,
such that any VSO ranking function $\rho_f$ for $\mathcal K$  (with DMO descent)
must use at least $2^n$ different vectors.
\end{thm}

\begin{fig1}{A partial view of two graphs from Construction {\thesection}.3.}{t}{fig-calK}
{\setlength{\unitlength}{0.46pt}\footnotesize
\begin{center}
\begin{picture}(200,240)(100,-55)
	      \put( 85,-20){\framebox(120,205){}}
             \put( 95,150){${x}_0$}
			 \put( 95,110){${x}_1$}
              \put( 95, 70){${y}_1$}
              \put( 95, 30){$x_n$}
              \put( 95, -10){$y_n$}
          
		\dashline{5}(85,130)(205,130)
        \dashline{5}(85, 50)(205, 50)

	      \thicklines
	      \put(115,153){\vector(1,0){50}}
	      \put(115,152){\vector(1,0){50}}
	      \thinlines
	      \put(115,33){\vector(1,0){50}}
	      \put(115,-7){\vector(1,0){50}}
	      
	      \put(115,73){\vector(4,3){50}}
	      \put(115,113){\vector(1,0){50}}
	      
             \put(175,150){${x}_0$}
              \put(175,110){${x}_1$}
              \put(175, 70){${y}_1$}
              \put(175, 30){${x}_n$}
              \put(175, -10){${y}_n$}
	      \put(120,-45){$H_x$}
\end{picture}
\hspace{0.3in}
\begin{picture}(200,240)(100,-55)
	      \put( 85,-20){\framebox(120,205){}}
             \put( 95,150){${x}_0$}
			 \put( 95,110){${x}_1$}
              \put( 95, 70){${y}_1$}
              \put( 95, 30){$x_n$}
              \put( 95, -10){$y_n$}
          
		\dashline{5}(85,130)(205,130)
        \dashline{5}(85, 50)(205, 50)

	      \thicklines
	      \put(115,73){\vector(1,0){50}}
	      \put(115,72){\vector(1,0){50}}
	      \put(115,113){\vector(4,-3){50}}
	      \put(115,112){\vector(4,-3){50}}	      \thinlines
	      \put(115,33){\vector(1,0){50}}
	      \put(115,-7){\vector(1,0){50}}

             \put(175,150){${x}_0$}
              \put(175,110){${x}_1$}
              \put(175, 70){${y}_1$}
              \put(175, 30){${x}_n$}
              \put(175, -10){${y}_n$}
	      \put(120,-45){$H_y$}
\end{picture}
\end{center}
}
%\vair
\end{fig1}

%%%%%%%%%%%%%%%%%%%%%%%%%%%%%

\subsubsection*{Construction {\thesection}.3: SCT instance $\mathcal K$}
{\abovedisplayskip=\abovedisplayshortskip
The variables in this example are named $x_0,\dots,x_n$ and $y_1,\dots,y_n$. 

Let ${\mathcal K} = \{G_1,\dots,G_{n-1},H_x,H_y\}$, where
the $G_k$ graphs are permutation graphs operating on pairs $(x_i,y_i)$:
$$\begin{array}{rcl}
G_k &=& \{ x_k \to x_{k+1}, x_{k+1}\to x_k, y_k \to y_{k+1}, y_{k+1}\to y_k \} \\
      &&  \cup \{ x_i\to x_i, y_i\to y_i \mid i\neq k,k+1\} \\
      &&  \cup \{ x_0 \stackrel{\downarrow}{\to} x_0 \} \\
\end{array}$$
Graphs $H_x$, $H_y$ operate specially on the first pair:
$$\begin{array}{rcl}
H_x &=& \{ x_1 \to x_1, y_1\to x_1 \}
         \cup \{ x_i \to x_i, y_i\to y_i \mid i>1 \} 
		 \cup \{ x_0 \stackrel{\downarrow}\to x_0 \}; \\
H_y &=& \{ x_1 \stackrel{\downarrow}\to y_1, y_1\stackrel{\downarrow}\to y_1 \}
         \cup \{ x_i \to x_i, y_i\to y_i \mid i>1 \}
		 \\
\end{array}$$\par}

\blem
$\mathcal K$ satisfies SCT.
\elem

\bprf
Variable $x_0$ guarantees descent, except for multipaths that contain $H_y$.
 It suffices to prove that for each such finite multipath $M= M_1H_yM_2$,
$M^\omega$ includes an infinitely-descending thread\footnote{
This sufficiency claim is implicit in the proof of Theorem~4 in~\cite{leejonesbenamram01}.%
}.
Note that $M$ has this
property if and only if $M' = M_2M_1H_y$ does. Now, $M'$ has
at least two descending threads that end at $y_1$,
beginning at $x_i$ and $y_i$ for some $i$. If $i=1$,
the thread from $y_1$ is multiplied infinitely in $(M')^\omega$. If $i\ne 1$,
observe that the effect of $M'$ on the pairs is
to permute them, so there is a
$k$ (the order of the permutation)
such that $(M')^k$ has a $y_1$ to $y_1$ thread as in the simple case, and we have
the same conclusion regarding $(M')^\omega$.
\eprf

In proving the lower bound on the number of vectors we use tools similar to those of
the previous proofs. Now, however, we restrict the orderings of variables
on which we focus
in the proof, so that the pairs are kept in increasing order, i.e.,
$$\max(x_1,y_1) < \min(x_2,y_2) < \max(x_2,y_2) < \min(x_3,y_3) < \cdots$$
and, as usual, $x_0$ is larger than the rest.  The relative order among the elements of each pair
may change.
The ordering of pairs is described by a function $\alpha:\{1,\dots,n\}\to\{{\tt x},{\tt y}\}$
that indicates, for every pair, which variable has the smaller value. Let $S'_n$ be the set of
such functions. Thus the order of values in a given state is described by an element
of $S'_n$, and we use $S'_n$ as the domain of $\rho_f^*$.

The basic
properties of a ranking function $\rho_f$ for this instance
follow the pattern of previous
examples, so we omit a detailed proof:
\begin{enumerate}[(1)]
\item
$x_0$ appears last.
\item
The constants in the even positions below $2n$ coincide for all
vectors returned by $\rho_f^*$.
%The constant in every even position below $2n$ is the same over all
%vectors returned by $\rho_f^*$.
\item
For every odd position, the size of the set is the same in all vectors returned by
$\rho_f^*$.
\end{enumerate}
The property specific to this construction is as follows.

\blem
Let $\alpha\in S'_n$.
Let $\vec v_\alpha = \rho_f^*(\alpha)$. Then from every pair $\{x_i, y_i\}$, only
the element selected by $\alpha$, that is, the variable of smaller value,
 is present in $\vec v_\alpha$.
\elem

\bprf
The proof is by contradiction, as usual.
Consider the first position that violates the lemma.
Suppose that it is an odd position $2i-1$, containing a set $A$.

\emph{Case 1}: $x_j\in A$, but $\alpha(j) = {\tt y}$.  Suppose first, for simplicity, that
$j=1$.
Consider a state $s$ described by $\alpha$, and
a state $s'$ where the value of $x_1$ is greater than
its value in $s$, the value of $y_1$ is smaller than in $s$, other values
do not change, and neither does the relative order among all variables. Thus,
the order at $s'$ is also given by $\alpha$. Note that $H_y\models s\mapsto s'$.
The multiset of values of $A$ increases from $s$ to $s'$, unless $A$ includes $y_1$.
So, we conclude that both $x_1$ and $y_1$ are in $A$. Now, consider $H_x$;
we can easily have a transition that increases the multiset value of $A$
and consequently a lexicographic increase in the value of $\rho_f$.

The case where $j>1$ is proved by the same argument, using $M_{\xi_{j1}} H_y M_{\xi_{j1}}$
(here $M_{\xi_{j1}}$ is a multipath that exchanges the pair $(x_j,y_j)$ with $(x_1,y_1)$).

\emph{Case 2}: $A$
includes $y_j$, while $\alpha(j) = {\tt x}$.
Again, for simplicity, assume $j=1$ (otherwise use $M_{\xi_{j1}}$ as above).
Consider a state $s$ described by $\alpha$,
and a state $s'$ where the value of $y_1$ is greater than
its value in $s$, other values (except for $x_0$)
do not change, and neither does the order. This agrees with $H_y$. Now,
$\rho_f(s') = s'(\vec v_\alpha)$;
the multiset of values of $A$ increases from $s$ to $s'$, a contradiction to
correctness of $\rho_f$.
\eprf

The proof of Theorem~\ref{thm-lb3} is now concluded since the  lemma implies
that a distinct vector corresponds to each of the $2^n$ orderings in $S'_n$.

Remark 1: $2^n$ is tight for the example; a possible ranking function for $\mathcal K$
is one that returns the minimum over a set of vectors, all of length 3,
constructed as follows%
\footnote{Technically, we need to add a dummy fourth element to match the
definition of $V^f$.}:
The first position contains a set of $n$ variables---%
one from every pair. All $2^n$ such sets are present. For a vector that
begins with set $S$, the next position contains the number of $y$'s in $S$.
The third position is $\{x_0\}$.

Proof: to prove lexicographic descent, the non-trivial case is a transition that
does not decrease the minimum among the multiset values in the first position 
(clearly, it never increases).
Suppose that in such a transition, the old minimum vector was 
$\vec v_1 = \langle S_1,k_1,\dots\rangle$
and the new one is $\vec v_2 = \langle S_2,k_2,\dots\rangle$.
If $k_2<k_1$, we have descent. So, assume $k_2\ge k_1$, and consider what transition
was taken.
\begin{enumerate}[$\bullet$]
\item
 If it is $H_x$, $S_2$ can only differ from $S_1$ if the minimum among
$\{x_1,y_1\}$ moved from $y_1$ to $x_1$ (the new value of $y_1$ being at least as large).
But then the number of $y$'s in $S_2$ is smaller. If $S_2$ and $S_1$ are the same,
we have lexicographic descent because $x_0$ decreases.

\item
 If it is $H_y$, $S_2$ must differ in value from $S_1$. In fact, there is dual multiset
descent in $\{x_1,y_1\}$.

\item
 If it is $G_i$, given that the multiset does not decrease,
 $S_2$ has to be the same set as $S_1$. So $k_2=k_1$, and $x_0$ decreases.\qed
\end{enumerate}

Remark 2: The lower bound dropped from $2^{\Theta(n\log n)}$ to $2^{\Theta(n)}$.
What is the true complexity? It's an intriguing possibility that the use of multisets 
(beyond SMO) might decrease the ranking-function complexity 
(for a single flow-point) to $2^{O(n)}$.

\subsection{On the program size of free-form ranking functions}
\label{sec-programsize}

A ``free form'' ranking function is described by a program; consider, for instance,
a program that sorts the variables and outputs the sorted list.
The size of a program expressing this function can be much smaller than
the number of vectors in its image ($n!$).
Giving an explicit lower bound on the size of general programs
seems quite hard. Instead, \cite{Ben-Amram:ranking} 
argues that polynomial ranking functions are
very unlikely to exist, under standard complexity-theoretic assumptions.
 This is shown for all-strict SCT,
using a proof that the decision problem (does an all-strict SCT instance terminate?)
is PSPACE-hard. 
We note that the decision problems for fan-in free (or fan-out free) SCT 
are also PSPACE-hard~(\cite{leejonesbenamram01} gives the
proof for fan-in free graphs and the fan-out free case follows easily).
Thus, the same complexity-theoretic conclusion
applies to these sub-problems as well.

\section{The Ordinal Height of Ranking Functions}

The fact that \emph{existence} of a global ranking function follows from
a termination proof of the local type is immediate from the fact that
every terminating program has a global ranking function. However what can
be deduced from the form or number of local ranking functions 
on the form or complexity of a global one is a difficult question.
A recent paper by Blass and Gurevich~\cite{BG:ranking} settles the
question of the \emph{ordinal height} of the global ranking function.
\begin{enumerate}[$\bullet$]
\item
Consider any transitive relation $R$ (in our setting, this will be the
transitive closure of the transition relation of the subject program,
or the SCT transition system as described in Definition~\ref{def-transitionsystem}). The goal of the
``termination proof'' is to establish that $R$ is well-founded.
\item
The goal is achieved by describing a finite set $U_1,\dots,U_m$
of well-founded relations such that $R\subset U_1\cup\dots\cup U_m$.
\item
Blass and Gurevich give a general upper bound on the ordinal height $|R|$
of $R$ in terms of the ordinals $\alpha_i = |U_i|$.
\end{enumerate}
In our setting, the covering relations $U_i$ will include two types of
relations:
\begin{enumerate}[(1)]
\item
For all pairs $(f,g)$ of distinct flow-points, there is a relation 
that includes all pairs of states $((f,\sigma),(g,\sigma'))$. This
relation is well-founded because there are actually no chains in it of
length greater than one. Removing these parts from $R$ leaves only the
part that describes cycles, i.e., pairs $((f,\sigma),(f,\sigma'))$.
\item
The second group of relations $U_i$ covers all cycles.
\end{enumerate}
Obviously, the latter group is the interesting one for termination proofs
as well as for the ordinal-height question, and to simplify the present
discussion, we can restrict our attention to the case $|F|=1$, so
that only the second kind is left.

In order to get the best bound
out of the Blass-Gurevich theorem, we want to cover $R$ in the most
economical way. To this end we use 
the following theorem, a rewording of~\cite[Corollary~1]{Codish-et-al:05}.

\bthm[Codish-Lagoon-Stuckey]
Size-change graph $G:f\to f$ over
parameters $x_1,\dots,x_n$ satisfies SCT if and only if
$$G \models (f,\sigma)\mapsto (g,\sigma') \Longrightarrow
     \bigvee_{1\le i\le n} \sigma(x_i) > \sigma'(x_i) \,.$$
\ethm

Simply put, this means that the set of $n$ relations $\sigma(x_i) > \sigma'(x_i)$ covers
the transition relation of $G$. Therefore, the Blass-Gurevich result
applies with exactly $n$ relations $U_i$, all isomorphic to $\val$.
In this simple case, Blass and Gurevich's bound becomes $\alpha^{n}$,
where $\alpha$ is the ordinal height of $\val$ (e.g., $\omega$ when $\val$
is the natural numbers, the most usual case).

Our explicit constructions match this bound, at least for typical
domains. Consider the construction where the range of the ranking function
is described by vectors of length $2n$, alternating parameters and
integers bounded by $B$. The order type of this range is $(B\alpha)^n$.
Whenever $\alpha$ is a multiple of $\omega$, this is exactly $\alpha^n$.

While the ordinal bound matches,
our results in this work are not a corollary of Blass and Gurevich's
result. Note that in the above argument
we covered all SCT instances with the same set of $n$ local functions.
Obviously, the identity of this set of functions cannot reveal any specific
structure of a given instance, as does an expression for a ranking function.

\section{Conclusion}

While deducing ranking expressions from size-change graphs has already
been shown possible before this work,
the constructions in this paper
are simpler and more transparent than previously known.
They improve the upper bound on the size of the ranking expression
and in fact achieve optimality, in a certain sense.

The constructions employ reductions of SCT instances to
instances of a subclass of SCT, and applies SCT to data of composite types
(tuples and sets).
We feel that this technique is interesting in itself.

To argue for optimality, we have introduced a class of expressions that
(in our opinion) captures all possible ranking functions for general SCT,
and a complexity measure (number of different vectors in the image) under
which we are able to prove lower bounds.

Several theoretical problems remain. For example:
 \begin{enumerate}[$\bullet$]
 \item
For fan-in/-out free graphs,
will the use of multisets and multiset ordering allow ranking functions
of size $2^n$ to be constructed? Or is our lower bound loose?  
 \item
What is the complexity of ranking functions for general SCT, and how to generate them?

We remark that our construction relied on fan-out freedom in the construction of the instance
$\mathcal R$ (Section~\ref{sec:basic-fof}). Furthermore, it is not hard to verify that the forms
of ranking functions given by our constructions do not suffice for certain SCT instances
which are not fan-in or fan-out free.  The strict SCT instance shown on Page 
\pageref{fig-ex-m05} is such an example. On the other hand, practically, there is evidence
that fan-in free graphs are common. For example,
in analysing a benchmark of SCT instances derived from Prolog programs~\cite{BL:2006},
we discovered that fan-in occurred rarely once size-change graphs have been ``cleaned up'' by
 removing arcs unnecessary for the termination proof.
 \item
How can the construction algorithm given in this paper be improved? Note that as presented,
it may require doubly exponential time and space, despite the fact that the size of the
final result is bounded by $m\cdot 2^O(n\log n)$.
 \end{enumerate}

\noindent Practically, the choice of an algorithm for ranking-function construction and its usage
are also challenging. It is well known that even algorithms that are worst-case exponential
sometimes work sufficiently well in many practical cases. It is quite possible that human-written
programs will not require ranking functions of high complexity.
In this work, these practical questions have not been studied, as our goal was to examine the
theoretical  problem first.
Recent work by Ben-Amram and
Codish~\cite{BC:08:TACAS} proposes to use a different class of ranking functions
which does not cover all SCT instances, but has polynomial expression size; and it turns out
to suffice for the benchmark that was tried. The compact representation of the ranking
functions relies on the use of sets of variables inside tuples, an insight gained from the 
work described in this paper.

Even if a ranking-function construction is provided, the practical goals 
mentioned---certified termination, proof carrying code and execution time analysis---all
require additional research and implementation work for their realization.

\paragraph{Acknowledgment.} The authors are indebted to the anonymous referees whose
comments were instrumental in improving this paper.

% \bibliographystyle{plain}
% \bibliography{sct}

\end{document}